\documentclass[useAMS,usenatbib]{mnras}
\usepackage[utf8]{inputenc}
\usepackage{amssymb}
\usepackage{amsmath,bm}
\usepackage{graphicx}
\usepackage{multirow}
\usepackage{makecell}
\usepackage{hyperref}
\usepackage{xcolor,soul}

\def\obj{\mbox{GRO\,J0422+32}}

\begin{document}
\title[Photometry of the X-ray nova GROJ0422+32]
{I-band photometry of the quiescent black hole X-ray nova GROJ0422+32=V518\,Per:
modelling of the orbital light curve and estimation of the black hole mass}

\author[A.\,M. Cherepashchuk et al.]
{A.\,M. Cherepashchuk,$^{1}$\thanks{E-mail: cherepashchuk@gmail.com}
T.\,S. Khruzina,$^{1}$
K.\,E. Atapin$^{1}$\\
$^1$ Sternberg Astronomical Institute, Moscow State University, Universitetsky pr., 13, Moscow, 119991, Russia\\}
\pagerange{\pageref{firstpage}--\pageref{lastpage}}
\pubyear{2024}
\date{Accepted 2024 June 4. Received 2024 June 3; in original form 2023 September 7}

\label{firstpage} 
\maketitle

\begin{abstract}
We present new photometric observations of the X-ray nova \obj\ (V518\,Per) carried out in the I$_c$ band over 14 nights in 2020 -- 2023.
We had to revise the orbital period of the system, the new value $P_{\rm orb}=5^{\rm h}04^{\rm m}35\fs50 \pm 0\fs04$ is about one minute shorter than the period by \cite{Webb2000} but close to the result reported earlier by \cite{Filippenko1995}. The obtained folded light curve has a regular shape with a clear ellipticity effect and signatures of a slight heating effect. The simulations of this light curve in terms of the model of an interacting binary system allowed us to estimate the orbital inclination $i = 33^\circ - 49^\circ$ of the system and derive masses of the black hole $M_x=(6.5\pm2.9)M_\odot$ and companion star $M_v=(0.47\pm0.21)M_\odot$. This range of the black hole masses overlaps with the known gap of (2--5)\,M$_\odot$ in the distribution of compact objects masses but mostly it lies above the upper boundary of this gap. To obtain more precise estimates one needs to know the degree of heating of the donor star, so synchronous X-ray/optical observations of this system are desirable.
\end{abstract}


\section{Introduction}
The X-ray nova GRO\,J0422+32\,=\,V518\,Per is interesting in that it is suspected to contain a black hole of a very small mass $M_x=(3.97\pm0.95) M_\odot$, with a companion star of $M_v=(0.46\pm0.31) M_\odot$ \citep{GelinoHarrison2003}. Such a back hole mass falls into the gap of (2--5)\,M$_\odot$ in the mass distribution of relativistic objects, the nature of which is still unclear (see, for example, \citealt{Ozel2010BHmassdistribution}). Therefore, an independent estimate of the black hole mass in this system based on new observational data and the modern mathematical model of an X-ray binary system (e.\,g. \citealt{Khruzina2011,Cherepashchuk2019A0620,Cherepashchuk2019KVUMa,Cherepashchuk2021})  looks very intriguing.

The system \obj\ consists of a low-mass M2V-type optical star filling its Roche lobe and an accreting black hole. It was discovered during an X-ray outburst by the {\it Compton Gamma Ray Observatory} on 5 August 1992 \citep{Paciesas1992IAUC}. The detection was confirmed by the {\it Mir-Kvant} and {\it Granat} space observatories \citep{Sunyaev1993}, which showed that the source spectrum is typical for the `low' state of X-ray binaries; this state is characterized by the almost complete absence of a soft spectral component and the presence of a hard power-law `tail' extending up to 1--2~MeV.

The X-ray outburst was accompanied by an optical one, the detection of the optical counterpart of $m_V\simeq 12.5^m$ and its spectroscopic confirmation were reported by \cite{Castro-Tirado1992IAUC} and \cite{Wagner1992IAUC}. Subsequent optical monitoring showed a very slow decrease of the source brightness and the presence of mini-outbursts during the next 18 months \citep{ChevalierIlovaisky1995,Goranskii1996}. When the system returned to a quiescent state ($m_V\approx22.4$, \citealt{Zhao1994IAUC}), it became possible to observe optical variations caused by the effect of ellipticity of the donor star, which allowed to measure the orbital period of the system $P_{\rm orb}=0.212265$ day \citep{ChevalierIlovaisky1994IAUC}.

The optical spectroscopy carried out in the quiescent state by \cite{OroszBailyn1995} and \cite{Casares1995} revealed spectral lines of the donor start (TiO bands), which allowed to classify it as an M-type dwarf. In the quiescent state, the star contributes about $(35\pm6)\%$ in the $\lambda6000$--$6500\AA$ range and $(52\pm8)\%$ in $\lambda6700$--$7500\AA$ to the total optical emission of the system \citep{Casares1995}, while the remaining part of the emission comes from the accretion disc and the region of interaction between the disc and incoming gas stream. The I-band light curve taken in December 1994 has an average magnitude  $\left<{m_I}\right>=20.22^{m}\pm 0.07^{m}$ \citep{Casares1995}, and its shape can be well explained by the ellipticity effect with a peak-to-peak amplitude of $\Delta m_I\simeq 0.1^{m}$. The observations obtained two months earlier by \cite{OroszBailyn1995} demonstrated $\left<{m_I}\right>=20.03^m\pm0.06^m$ with an amplitude $\Delta m_I\simeq 0.15^m$. Analysing the light curve taking into account ellipsoidal modulations and contribution from the accretion disc, \cite{Casares1995} have constrained the orbital inclination within $i=30^\circ- 35^\circ$ and, assuming the mass of the M2V donor $M_v=0.4 M_\odot$, estimated the mass of the compact object as $M_x=(2.5 - 5.0)$\,$M_\odot$. 

The first high-quality spectra of \obj\ were presented by \cite*{Filippenko1995}, who carried out 21 observations with the W.\,M. Keck 10-m telescope in 1994--1995. Studying absorption lines of the normal star, the authors constructed its radial velocity curve (a semi-amplitude is $K_v=380.6\pm6.5$~km/s) and measured the orbital period $0\fd21159\pm0\fd00057$; the derived mass function is $f_v(M_x)=1.21\pm0.06 M_\odot$. In order to track the motion of the compact object, they analysed the behaviour of the H$\alpha$ emission line, which, however, demonstrated a double-peaked profile and had shown strong variations in the 1994 data and moderate in the 1995. The fitting of a Gaussian profile to the line wings gave a radial velocity curve with $K_x=41.6\pm3.2$~km/s. The authors note that this velocity curve appeared to be shifted by 253\degr\ respect to the donor's one (instead of 180\degr), which may indicate that the obtained curve does not necessary reflect the true motion of the compact object. Under the assumption that it does, \cite{Filippenko1995} derived the mass ratio $Q=M_v/M_x=0.1093\pm0.0086$ ($q=M_x/M_v\simeq9.15)$ and the mass of the compact object $M_x=3.57\pm0.34 M_\odot$ (for an M2-dwarf donor of $0.39\pm0.02 M_\odot$). The same spectra were later reanalysed by \cite{Harlaftis1999Keck}. They measured the value $V_{\rm rot}\sin{i}=90_{-27}^{+22}$~km/s ($1\sigma$ confidence interval) from rotational broadening of absorption lines of the donor star and estimated the mass ratio as $Q=0.116_{-0.071}^{+0.079}$ ($q\simeq8.62$). According to \cite{Harlaftis1999Keck}, the companion star (of a spectral type M2$_{-1}^{+2}$) provides ($61\pm4$)\% of the total optical luminosity of the system at red wavelengths.

%\cite{Harlaftis1999Keck} reanalysed the observations taken by \cite{Filippenko1995} and estimated the value $v\sin{i}=90_{-27}^{+22}$~km/s ($1\sigma$ confidence interval) from the rotational broadening of absorption lines of the donor star. This yields the mass ratio $Q=0.116_{-0.071}^{+0.079}$ ($q\simeq8.62$).  According to \cite{Harlaftis1999Keck}, the companion star (of a spectral type M2$_{-1}^{+2}$) provides ($61\pm4$)\% of the total optical luminosity of the system in the red part of the spectrum.
%\hide{In contrast to many other  X-ray novae, the \obj\ companion star does not exhibit the Li $\lambda6708\AA$ absorption line.}

\cite*{Kato1995} presented photometric observations carried out a few months after the outburst. They discovered the period $0\fd2157\pm0\fd0010$ that is supposed to be associated with the flux modulations produced by tidal deformations of the accretion disc by the secondary star~--- the so-called 'superhump' period known in some other X-ray transients. It can be observed in photometric data in the bright state when the optical emission is dominated by the accretion disc. Comparing this period with the orbital one, %\hide{$0\fd21211\pm0\fd00002$}
\cite{Kato1995} gave an independent estimate for the mass ratio $q \simeq 5.4 - 20.5$. 

\cite{Garcia1996} analysed spectroscopy taken in October 1993, when the system was about 2\,mag above quiescence. They were unable to detect narrow absorption lines of the companion star but they constructed the radial velocity curve of the compact object fitting the wings of the double-peaked H$\alpha$ line. The obtained semi-amplitude $K_x=34\pm6$~km/s corresponds to the mass ratio $Q\simeq0.09$ ($M_x\simeq 5.6 M_\odot$). The I-band photometry obtained by the same authors in quiescence \citep{Callanan1996} showed a very small ellipsoidal modulation with an amplitude $\sim 0.06^{m}$, from which they concluded that the orbital inclination of the system must be $i \leq 45^\circ$. Also they noted that the companion star luminosity implies a distance of $\lesssim 2.2$~kpc, so the X-ray luminosity of the system in quiescence has to be below $2\times10^{32}$~egs~s$^{-1}$. Actually, this upper limit was established based on observations with \textit{ROSAT}, which had not detected the source. More accurate flux measurements carried out with \textit{Chandra} yielded a luminosity in the quiescent state of $L_X\simeq 8\times 10^{30}$~erg~s$^{-1}$ (0.5--10.0~keV, \citealt{Garcia2001}).

\cite{ChevalierIlovaisky1996} obtained photometry in the I$_c$ and R$_c$ filters in the quiescent state (from November 1994 to February 1995). Their R-band light curve is a double wave ($\Delta m_R \simeq 0.2^{m}$) with a variable shape and variable peak heights. In some nights the object brightness in the R$_c$ band was almost constant within $\pm0.05^m$. Sometimes there were some hints to the heating effect.

\cite{Webb2000} performed 53 simultaneous spectroscopic ($\lambda6900-9500\AA$) and photometric (I-band) observations which caught the system in a very low state with $m_I=20.44\pm0.08^{m}$ (December 1997).
The authors built a radial velocity curve of the companion star ($K_v=378\pm16$~km/s) based on the TiO bands ($\lambda7055\AA$ and $\lambda7589\AA$) in its spectrum and obtained a new estimate for the mass function $f_v(M_x)=1.191\pm0.021 M_\odot$. Also, using, among other, the data previously published by other authors, they have clarified the orbital period $P_{\rm orb}=0\fd2121600\pm0\fd0000002$. The system inclination derived from the amplitude of ellipsoidal variations is $i<45^\circ$ (the optical star contribution is $(38\pm2)\%$ in the $\lambda6950-8400\AA$ range). The companion spectral class is re-estimated as M4--5 which implies a distance of $1.39\pm0.15$~kpc. The I-band light curve ($\Delta m_I\simeq0.14^m$) shows almost no signs of the secondary minimum (significant heating effect).

%In the paper by \cite{Webb2000} the simultaneous spectroscopic ($\lambda6900-9500\AA$)
%\hide{the 4.2-m William Herschel telescope})
%and photometric (I-band)
%\hide{the 2.5-m Isaac Newton telescope})
%observations were performed. The object was in the lowest state with $m_I=20.44\pm0.08^{m}$. Based on the TiO bands ($\lambda7055\AA$ and $\lambda7589\AA$), the authors clarified the radial velocity semi-amplitude of the optical star $K_v=378\pm16$~km/s and obtained a new estimate for its mass function $f_v(M_x)=1.191\pm0.021 M_\odot$. Also they provided a more precise value of the orbital period $P_{\rm orb}=0\fd2121600\pm0\fd0000002$ and, from the analysis of the ellipsoidal variability, estimated the system inclination as $i<45^\circ$ (the optical star contribution is $(38\pm2)\%$ in the $\lambda6950-8400\AA$ range). The companion spectral class is re-estimated as M4--5 which implies a distance of $1.39\pm0.15$~kpc. The I-band light curve ($\Delta m_I\simeq0.14^m$) shows almost no signs of secondary minima (the heating effect).

The first infrared photometry of \obj\ was presented by \cite{Beekman1997}. Analysing it together with optical data in the R$_c$ and I$_c$ filters, they estimated $i$ as between $10^\circ$ and $26^\circ$, which implies a much more massive black hole $M_x>9 M_\odot$. The obtained inclination is so low because the authors reject the possibility that the dips in the object's light curves seen during optical outbursts may be the signatures of partial eclipses as it has been proposed by some other authors; \cite{Beekman1997} insist that these dips are just random luminosity variations of the accretion disc.

%\cite{Beekman1997} have obtained the first infrared photometry of \obj. Analysing it together with the optical data in the R$_c$ and I$_c$ filters they estimated $i$ as between $10^\circ$ and $26^\circ$ which implies a much more massive black hole $M_x>9 M_\odot$. The obtained inclination is so low because \cite{Beekman1997} rejects the possibility that the dips in object light curves seen during outbursts may be the signatures of partial eclipses as it has been proposed by some other authors, suggesting that the dips are just random luminosity variations of the accretion disc.

\cite{GelinoHarrison2003} have published the results of the detailed J,\,H,\,K photometry that had been being performed for 37 months, from January 2000 to February 2003. This IR light curve exhibits clear ellipsoidal modulation ($\Delta m_J=0.17^m$, $\Delta m_K=0.14^m$) without changes of their shape or mean level, which allows to measure the system inclination with a high accuracy $i=45^\circ\pm2^\circ$ and clarify the masses of the binary components: $M_x=(3.97\pm0.95) M_\odot$ and $M_v=(0.46\pm0.31) M_\odot$. Thus, the authors suggest that \obj\ contains a black hole with one of the lowest mass ever reported, which has to fall into the 2--5\,M$_\odot$ gap in the mass distribution of compact objects. 

%\chk{On the other hand, Beekman et al. (1997) who also analysed IR photometry} along with optical data in the R$_c$ and I$_c$ filters, obtained a much different estimate of the orbital inclination $i=13^\circ\pm31^\circ$ \hide{(for the disk contribution of 30\% and ellipsoidal variation of $\Delta m_R\simeq0.08^m$)} which implies a much more massive black hole $M_X>9 M_\odot$. The obtained inclination is so low because \cite{Beekman1997} rejects the possibility that the dips in the object light curve seen during outbursts may be signatures of partial eclipses as it has been proposed by some other authors, suggesting that the dips are just random luminosity variations of the accretion disc.

\cite{Reynolds2007Keck} have presented the K-band photometry performed in superior atmospheric conditions with the 10-m W.\,M.~Keck telescope in November 1997.
Their light curve, obtained when the system was in quiescence, has shown negligible elipsoidal variations ($\Delta m_K \leqslant 0.1^m$) but high flickering $\sim 0.15^m$. The authors pointed out that previous attempts to constrain the black hole mass in \obj\ based on IR-photometry are subject to significant uncertainty and considered three mechanisms that can potentially produce the observed flux variations in this band instead of ellipsoidal modulations: spots on the M-type donor, the flux contamination originating in a radio jet and variability of the accretion (advection) disc. They note that isolated rapidly rotating M-stars (with $v_{\rm rot} \sin i\approx 60 - 70$ km/s, e.\,g. HK\,Aqr and RE\,1816+541) show quasi-periodic variability with an amplitude of $\sim0.09^m$ (V band).

In view of the above-described discrepancies in estimate of the black hole mass in \obj, as well as due to the complexity of photometric behaviour of this system, the accumulation of new observational data and independent estimation of the orbital inclination and black hole mass are of great interest. In this paper we present new I-band photometry of the system and report our results of the detailed modelling of the system's light curves which allowed us to estimate the orbital inclination and apply constraints to the black hole mass.

\begin{figure}
\includegraphics[width=0.48\textwidth]{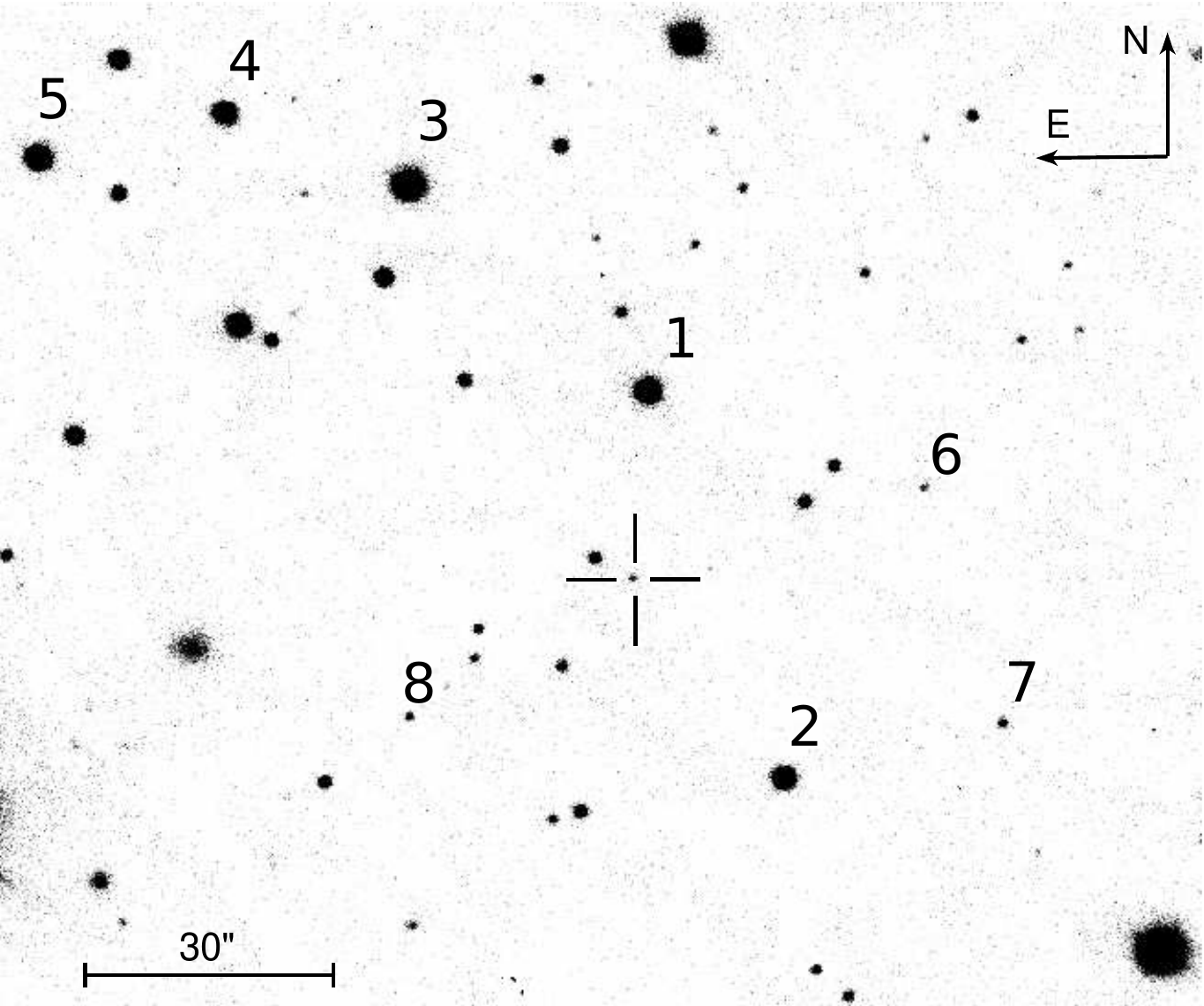}
\caption{Environment of \obj, the object is marked by a cross-hair.}
\label{fig1}
\end{figure}

\section{Observational data}
The object was observed with the 2.5-m telescope \citep{Potanin2017,Shatsky2020gbar} of Caucasus Mountain Observatory of SAI MSU (Sternberg Astronomical Institute, Lomonosov Moscow State University) on 14 occasions from March 2020 to January 2023. Observations were conducted in series of 2--5 hours long; all the images were taken in the I$_c$ filter ($\lambda=883$~nm, $\Delta \lambda=310$~nm) and, in the most cases, had an exposure time of 300~s. The detector was the NBI 4k$\times$4k camera manufactured by N.\,Bohr Institute, Copenhagen, which is a mosaic of two EEV\,4482 back illuminated CCD chips. The field of view is ${10'\times 10'}$, the image scale is 0.15''/pix. The available 14 data sets do not fill the entire period of observations evenly, but form five compact groups (hereafter, `seasons') separated from each other by months. The details are shown in Table~\ref{tab1:obs}.

In all the cases, the object and its surrounding stars, some of which were used as comparison and control stars (Fig.~\ref{fig1}), were placed in the chip area where fringes are negligible (the $3.5'\times 5'$ region at the central part of the left chip). As an additional but maybe unnecessary precaution, we shifted each frame respect to the previous by 3--5''. This was intended to convert possible systematic errors caused by fringes into random ones. The primary data reduction was carried out in a standard way including bias subtraction, flat-fielding and correction for the detector's non-linearity. Then each frame was tied to the world coordinate system using the \textsc{astrometry.net} package \citep{Lang2010}. Aperture photometry was performed with \textsc{iraf}. The aperture radius was 4--8 pixels depending on the seeing; the background was measured in an annular aperture around the object. 

\begin{table}
%\captionstyle{flushleft}
\caption{Observation log}
%\bigskip
\setlength{\tabcolsep}{4pt}
\begin{tabular}{cccrcc}
\hline\hline
S$^a$& JD$^b$ & Phase$^c$ & N$^d$ & $\left<m\right>^e$ & $\Delta m^f$\\ \hline
$s$1     & 8918.22$-$.30  & 0.728$-$1.106  &{\bf 23} & 20.334$\pm$0.018 & +0.018 \\
$s$2     &                &              & {\bf 114} & 20.292$\pm$0.007 & $-0.024$ \\
& 9108.40$-$.58  & 0.830$-$1.681  &42 &  &  \\
& 9109.43$-$.58  & 0.700$-$1.409  &37 &  &  \\
& 9110.43$-$.55  & 0.428$-$0.995  &23 &  &  \\
& 9112.37$-$.56  & 0.599$-$1.497  &12 &  &  \\
$s$3     & 9188.40$-$.54  & 0.042$-$0.704  &{\bf 42} & 20.208$\pm$0.016 & $-0.108$ \\
$s$4     &              &            & {\bf 123} & 20.361$\pm$0.011 & +0.045 \\
& 9276.17$-$.33  & 0.987$-$1.743  &41 &  & \\
& 9278.21$-$.35  & 0.631$-$1.293  &36 &  & \\
& 9279.21$-$.34  & 0.359$-$0.973  &24 &  & \\
& 9286.19$-$.30  & 0.358$-$0.878  &22 &  & \\
$s$5     &              &            & {\bf 154}     &
20.236$\pm$0.006 &  $-0.080$ \\
& 9944.35$-$.50 & 0.902$-$1.611   & 39 &  & \\
& 9945.31$-$.48 & 0.440$-$1.244   & 43 &  & \\
& 9957.39$-$.47 & 0.550$-$0.928   & 21 &  & \\
& 9960.21$-$.43 & 0.882$-$1.922   & 51 &  & \\
\hline
All & 8918.22$-$~~~~ & & {\bf 456} & 20.316$\pm$0.007 & \\
&~~~~$-$9960.43 & &  &  &  \\ \hline

\end{tabular}
{\it Notes.} $^a$Season number. $^b$Julian dates (2\,450\,000 subtracted) of the start and the end of the data set with an accuracy of two decimal places. $^c$Orbital phases calculated using the ephemerides eq.~(\ref{eq:eph_new}). $^d$Number of frames in the data set, values in bold indicate the total number of frames in the season. $^e$Mean I-band magnitude of the season. $^f$Difference from the the total mean, $\Delta m = \left<m_{\rm sj}\right> - \left<m_{\rm All}\right>$. 
\label{tab1:obs}
\end{table}

\begin{figure*}
\includegraphics[width=0.33\textwidth]{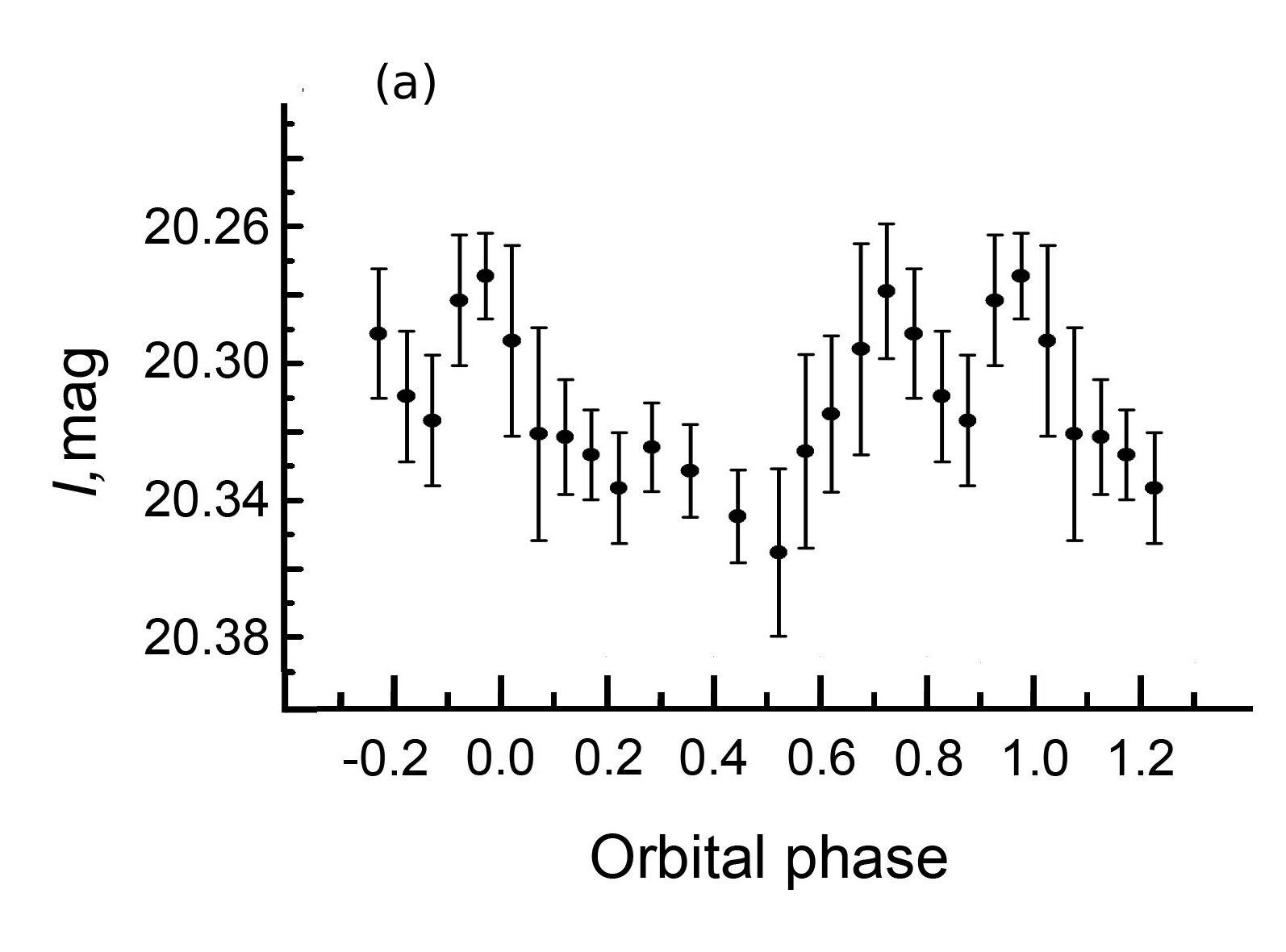}
\hspace{2pt}
\includegraphics[width=0.33\textwidth]{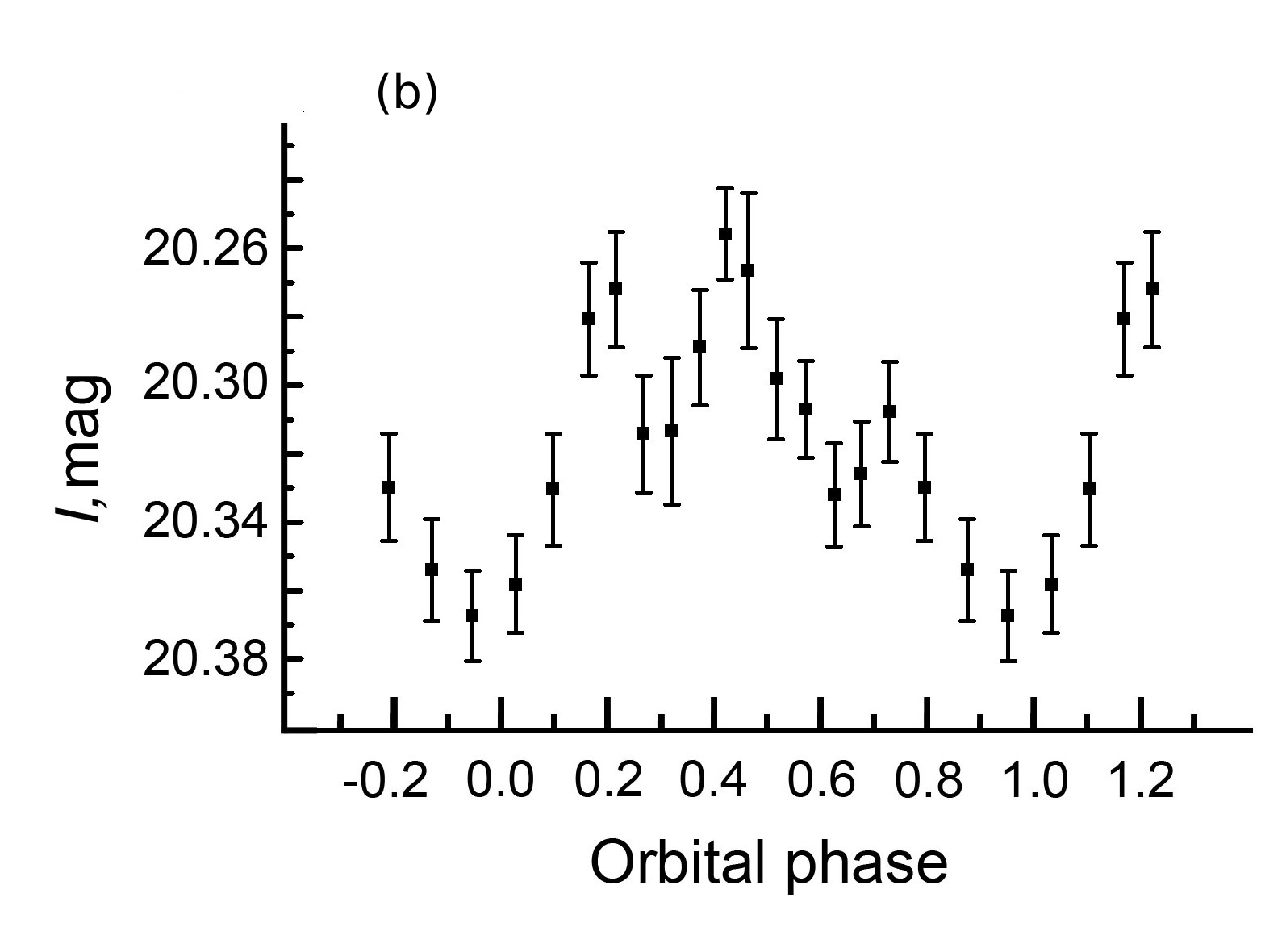}
\hspace{2pt}
\includegraphics[width=0.31\textwidth]{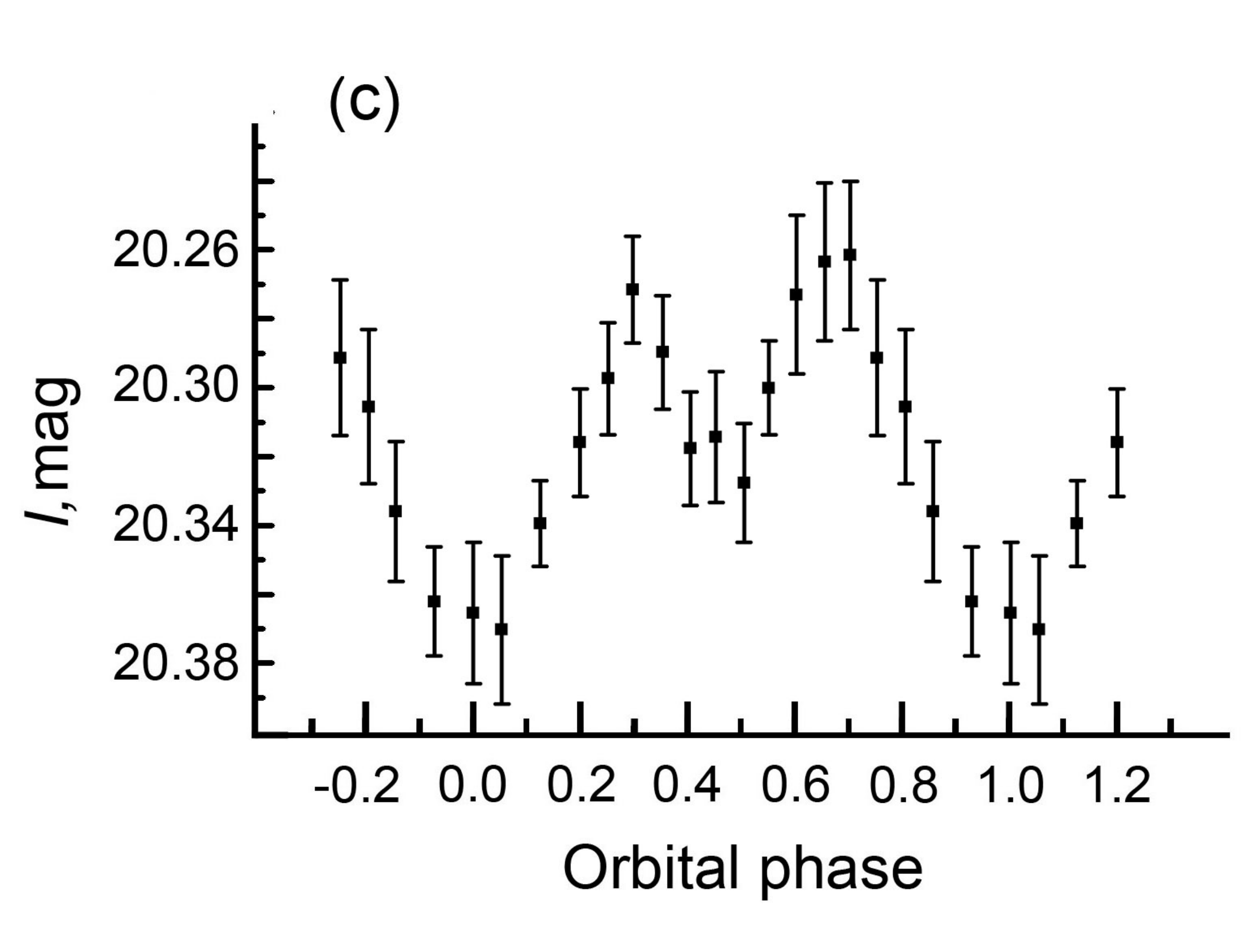}
\caption{Binned orbital light curves. {\it Left panel}: the orbital light curve obtained using the ephemeris from \protect\cite{Webb2000}. Its main minimum is located at $\varphi \simeq 0.5$ instead of $\varphi =0$ which suggests that the ephemeris needs to be refined. {\it Middle panel}: the light curve folded with a period corrected by $\approx0\fs24$ respect to that of \protect\cite{Webb2000} to compensate the phase offset. {\it Right panel}: the orbital light curve obtained with the ephemeris~(\ref{eq:eph_new}) and used in the modelling below.}
\label{fig:lcurve}
\end{figure*}

\begin{figure}
\includegraphics[width=0.48\textwidth]{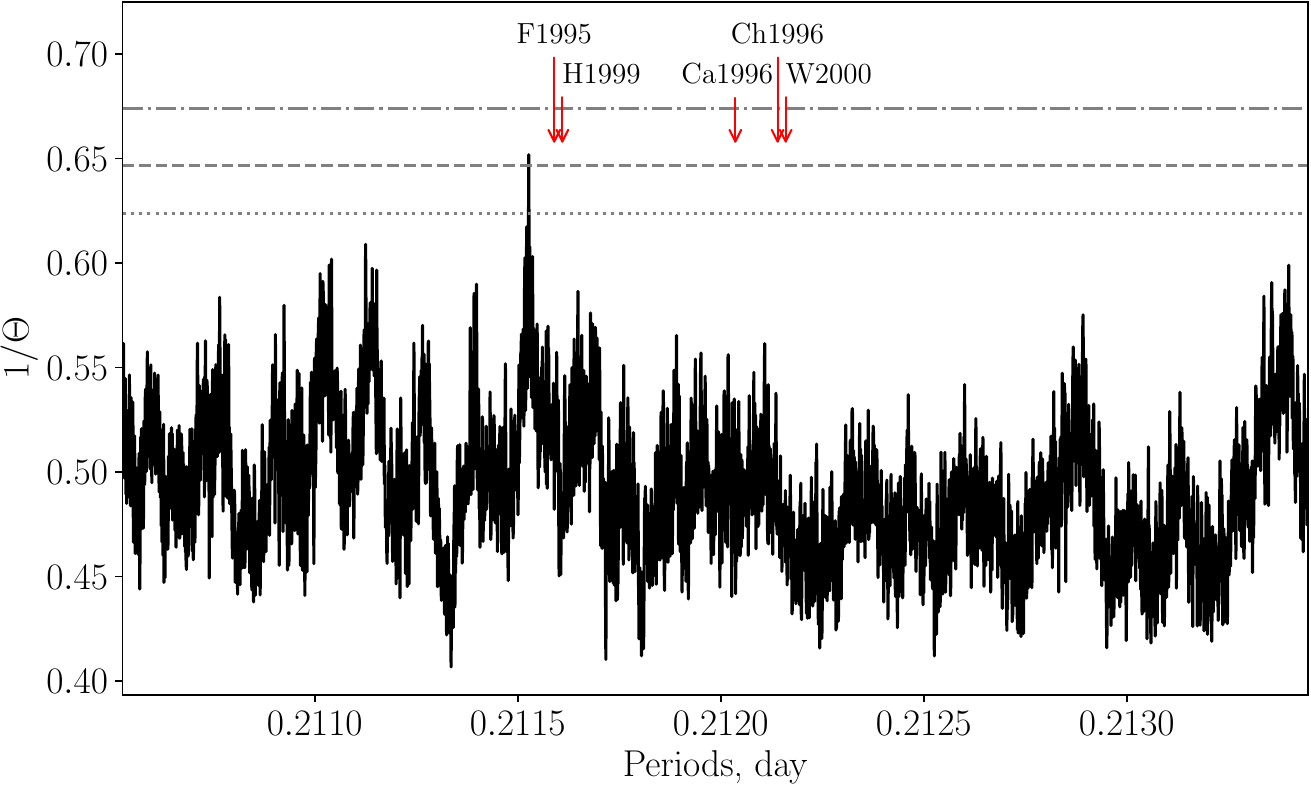}
\hspace{2pt}
\caption{Periodogram (the reciprocal of the Lafler-Kinman statistics $\Theta$) used to refine the orbital period of \obj. The highest peak corresponds to a period of $5^{\rm h}04^{\rm m}35\fs5$. Red arrows indicate the orbital periods reported by other authors: \protect\cite{Filippenko1995,Callanan1996,ChevalierIlovaisky1996,Harlaftis1999Keck,Webb2000}. Dotted, dashed and dash-dotted lines indicate $1\sigma$, $2\sigma$ and $3\sigma$ significance levels, respectively.}
\label{fig:fig03_periodogram}
\end{figure}

\begin{figure*}
\includegraphics[width=0.98\textwidth]{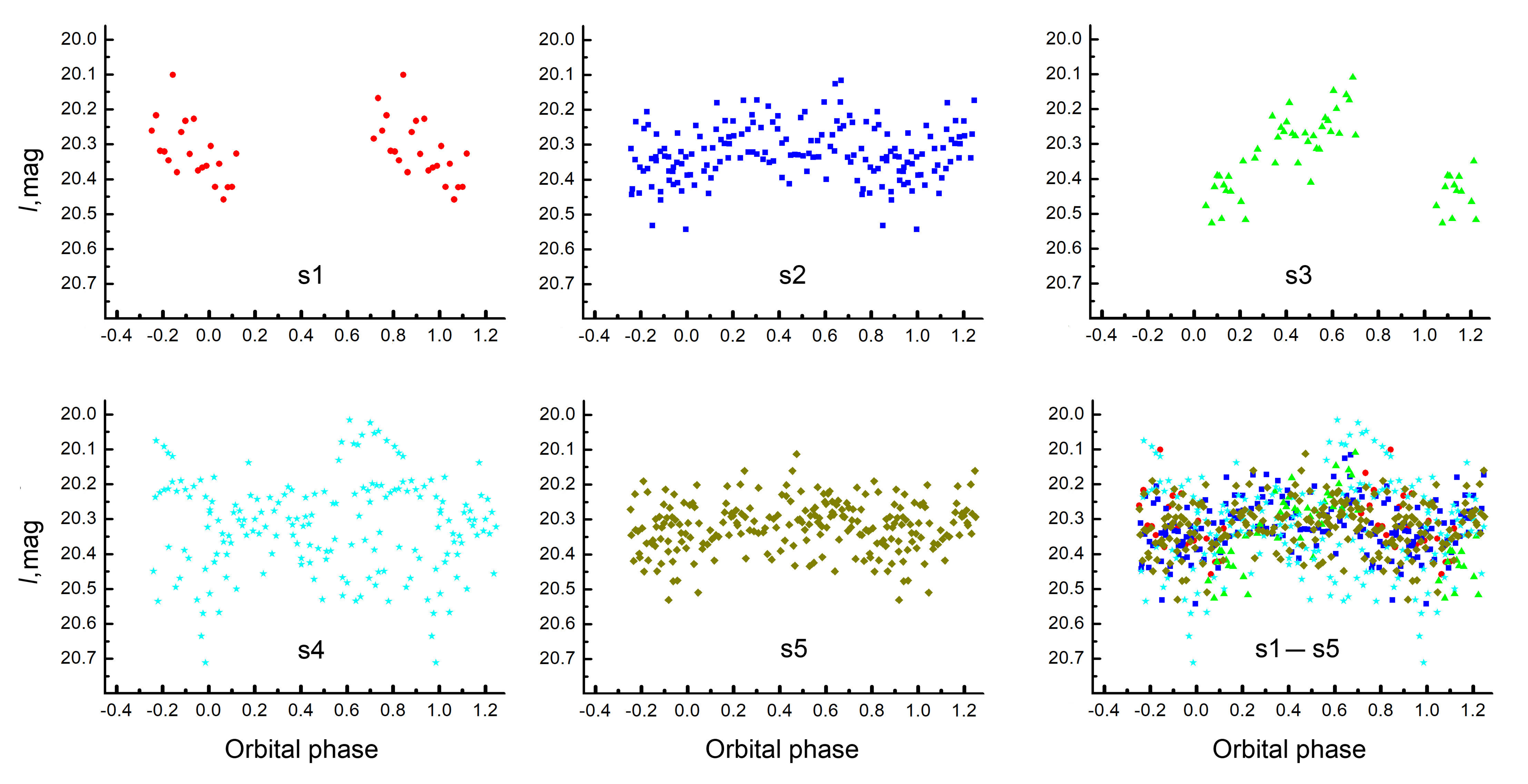}
\caption{Individual seasons folded with the ephemeris (\ref{eq:eph_new}). The label refer to the first column of Table~\ref{tab1:obs}.}
\label{fig:lcurve_seasons}
\end{figure*}

The \obj\ surroundings are shown in Fig.~\ref{fig1}. As a comparison star, we used the star \#1, whose I$_c$ magnitude was assumed to be equal to 15.50$^m$. This value was obtained based on the star brightness in the SDSS data; to convert the star magnitude from the SDSS photometric system to the Johnson-Cousins one we used calibrations from \cite{Jordi2006SDSS}. Stability of the comparison star across the seasons has been checked using the control stars \#\#2--5. We found that the season average magnitudes of each of these stars vary at most by 0.008$^m$ between the seasons, the standard deviations of individual magnitude measurements of the control stars are $\approx0.006^m$. This stability is sufficient, because the peak-to-peak amplitude of the \obj\ orbital variations is about 0.1$^m$.
The remaining stars \#\#6--8 of 19$^m$--20$^m$ were used for estimating the photometric accuracy for sources as weak as the studied binary system. We obtained that the magnitudes of these stars measured from individual 300-s. frames have standard deviations $\simeq 0.1^m$ (the formal photometric error provided by \textsc{iraf} is about twice as low).

As was mentioned in the Introduction, the system \obj\ shows notable variability (see, for example, \citealt{Reynolds2007Keck}) and can change its brightness between different epochs, which makes it difficult to construct a regular average orbital light curve. In particular, we found that, even after rejection of outliers, the object magnitudes in individual seasons remained slightly different from each other and from the mean level. We have accounted for these differences during constructing the average orbital curve, the applied corrections ($\Delta m$) are listed in the last column of Table~\ref{tab1:obs}. The object magnitude averaged over all the good observations (456 data points) is $\left<m_I\right>=20.316^m\pm0.007^m$.

%We found that individual seasons have slightly different mean levels, so in order to obtain the average orbital light curve, we have compensated for these differences using the $\Delta m$ values listed in the last column of Table~\ref{table1}.

In order to produce the orbital (folded) light curve, we initially tried to convolve our data with the orbital ephemeris from \cite{Webb2000}, which is the most up-to-date and have the smallest formal uncertainty of the orbital period $P_{\rm orb} = 5^{\rm h}05^{\rm m}30\fs624 \pm 0\fs017$:

\begin{equation}
{\rm Min}(\varphi=0)={\rm HJD\,} 245\,0274.4156(9) +0\fd2121600(2)\cdot E,
\end{equation}
here HJD denotes to the Heliocentric Julian Date, the digits in parentheses indicate the uncertainties related to digits at the last decimal place and the phase $\varphi= 0$ refers to the main minimum which occurs when the system is in the inferior conjunction (i.\,e. the companion star is between the observer and the compact object). However, the obtained orbital light curve has shown an offset of its main minimum to $\varphi\simeq0.5$ (Fig.~\ref{fig:lcurve}a). This indicates that either the value of the orbital period determined more than 20 years ago was not accurate enough or the period of the system has evolved with time. Assuming that the first option takes place, we calculated the minimal correction to the period required to eliminate the phase offset: $\Delta P\approx 2\times 10^{-6}$~day (the epoch was left unchanged). Then we convolved our data with some trial periods around $P+\Delta P$ and chose the one that gave the right position of the main minimum and the smallest scatter of data points. It yielded the ephemeris:
\begin{align}
\begin{split}
{\rm Min}(\varphi=0) &={\rm HJD\,} 245\,0274.4156(9) + 0\fd21216275\cdot E = \\
&= {\rm HJD\,} 245\,8918.1381978(9)+0\fd21216275\cdot E.
\end{split}
\label{eq:eph_near_webb}
\end{align}

The corrected period $5^{\rm h}05^{\rm m}30\fs862$ differs from that of \cite{Webb2000} by $\approx$\,$0\fs24$ which is enough to compensate for the phase offset but it produces a light curve with a quite unusual shape (Fig.~\ref{fig:lcurve}b): its secondary minimum is much shallower than the main one and is shifted to the phase $\varphi\cong0.3$ (instead of $\varphi=0.5$). In principle, this behaviour can be related to significant X-ray heating in the system. After the outburst, \obj\ was firmly observed in X-rays only once, during the observational campaign undertaken by \cite{Garcia2001} for the purpose of  measuring X-ray luminosities of black hole binaries in quiescence.
That observation gave $L_x\simeq 8\times10^{30}$~erg~s$^{-1}$ ({\it Chandra}, 0.5--10~keV), an order of magnitude lower than typical luminosities of M2V stars, which implies that heating of the donor star should be negligible. On the other hand, the X-ray luminosity of the system at the time of our optical observations is unknown, and some physical variability can still take place even the quiescent state. A similar effect was seen in A0620-00 whose orbital light curves were found to exhibit variations of their mean levels by $0.2^m$ and $0.3^m$ in the optical and IR ranges, respectively, on a time scale of a few months \citep{Cherepashchuk2019A0620}. 

Nevertheless, we have tried to find another $P_{\rm orb}$ that would be able to produce a folded light curve conforming the expectations for a low-mass X-ray binary in quiescence better. We used the Lafler-Kinman method (\citealt{LaflerKinman1965}, the WinEfk\footnote{\url{http://www.vgoranskij.net/software/}} software developed by V.\,P.\,Goranskij) which minimizes the squared difference in each pair of flux measurements with adjacent phases. The inverse Lafler-Kinman statistic as a function of orbital periods is plotted in Fig.~\ref{fig:fig03_periodogram}. The highest peak in the periodogram has a significance of 0.029\footnote{To estimate the significance, we simulated 10\,000 data sets via bootstrap resampling of the observed magnitudes, calculated a periodogram for each of them and built an empirical distribution of the heights of the highest peaks.} ($\approx$\,$2.2\sigma$) and corresponds to the period $5^{\rm h}04^{\rm m}35\fs50 \pm 0\fs04$. This orbital period is shorter by about a minute than the value by \cite{Webb2000} but it is consistent with the periods $P_{\rm orb} = 5^{\rm h}04^{\rm m}41^{\rm s} \pm 49^{\rm s}$ obtained by \cite{Filippenko1995} and $5^{\rm h}04^{\rm m}43^{\rm s} \pm 29^{\rm s}$ by \cite{Harlaftis1999Keck}. The corresponding ephemeris ( referring to the main minimum when the donor star is between the observer and the compact object) is 
\begin{align}
{\rm Min}(\varphi=0) = {\rm HJD\,} 245\,8915.5278+0\fd2115220(4)\cdot E.
\label{eq:eph_new}
\end{align}

The binned orbital light curve is shown in Fig.~\ref{fig:lcurve}c. It has the canonical shape of a double wave with a clear secondary minimum at $\varphi=0.5$. The ratio between depths of the two minima indicates modest heating of the donor star by X-rays. The light curve demonstrates slight asymmetry: the first hump of the double wave is a little lower than the second one, but a similar light curve morphology was observed earlier (Fig.~5c in \citealt{ChevalierIlovaisky1996}). These facts convinced us that it is this light curve (Fig.~\ref{fig:lcurve}c) that represents the true orbital variability of the system and should be used for the modelling. In other matters, simulations for the less probable orbital period described by the ephemeris~(\ref{eq:eph_near_webb}) has also been performed, we placed its consideration in Appendix.

Fig.~\ref{fig:lcurve_seasons} illustrates contributions of the individual seasons to the observed orbital variability of the system. Table~\ref{tab2:binned} lists the data values of the binned light curve (Fig.~\ref{fig:lcurve}c). As can be seen from the table, each phase bin includes from 14 to 52 individual measurements, the mean uncertainty is $0.019^m$, the full amplitude of the orbital variations is $0.109^m\pm 0.043^m$.

\begin{table}
\caption{Binned orbital light curve for the ephemeris (\ref{eq:eph_new}).}
\centering
\begin{tabular}{cccc}
\hline\hline
Phase & Magnitude & Error &~ $N$ \\
\hline
0.0011 & 20.365 & 0.021 & 30 \\
0.0541 & 20.370 & 0.022 & 22 \\
0.1249 & 20.339 & 0.013 & 45 \\
0.1996 & 20.316 & 0.017 & 20 \\
0.2515 & 20.297 & 0.016 & 21 \\
0.2976 & 20.271 & 0.016 & 14 \\
0.3532 & 20.289 & 0.017 & 22 \\
0.4046 & 20.317 & 0.017 & 24 \\
0.4525 & 20.314 & 0.019 & 22 \\
0.5049 & 20.327 & 0.017 & 22 \\
0.5511 & 20.300 & 0.014 & 24 \\
0.6030 & 20.273 & 0.023 & 27 \\
0.6548 & 20.263 & 0.023 & 26 \\
0.7027 & 20.261 & 0.022 & 26 \\
0.7526 & 20.291 & 0.023 & 26 \\
0.8056 & 20.305 & 0.022 & 24 \\
0.8567 & 20.336 & 0.020 & 28 \\
0.9284 & 20.362 & 0.016 & 52 \\
\hline
\end{tabular}
\label{tab2:binned}
\end{table}

\begin{figure}
\includegraphics[width=0.46\textwidth]{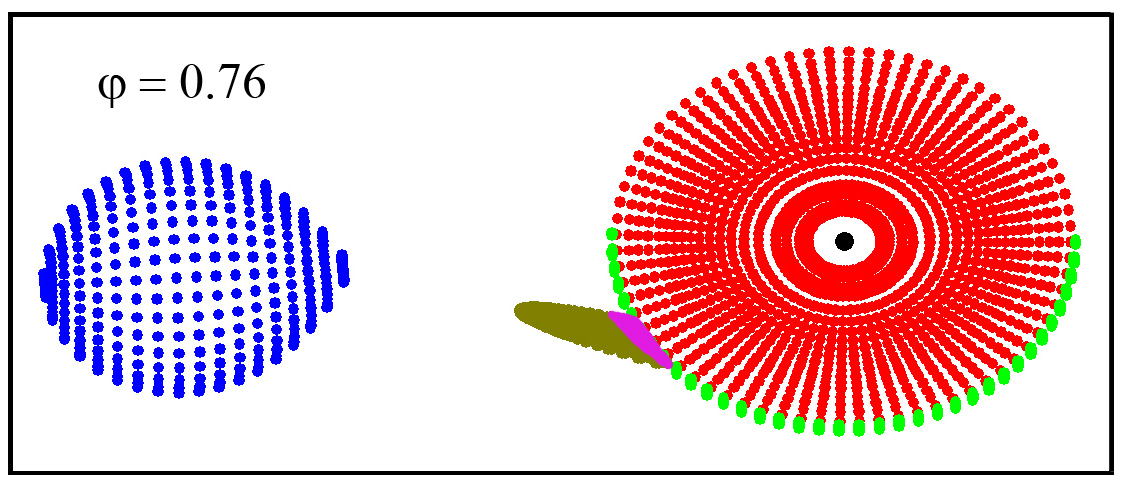}
\caption{Schematic representation of the \obj\ system for the mass ratio $q=14$ and orbital inclination $i=35^\circ$ (the remaining model parameters see in Table~\ref{tab3:model}) as it could be observed at the orbital phase $\varphi=0.76$. Three model components contributing to the optical emission are seen in the figure: the donor star (blue), the accretion disc (red) and the hot line (swamp green). Bright green marks the projection of the disc side surface onto the picture plane, purple marks the host spot. The binary is rotating counter-clockwise.}
\label{fig:scheme}
\end{figure}

\section{Modeling of the orbital light curve of \obj}
\subsection{Description of the model}
\label{sec:model_intro}
Simulation of the mean orbital light curve of \obj\ was carried out within the model of an interacting binary system (for details, see \citealt{Khruzina2011,Cherepashchuk2019A0620}). The modelling was performed under the following assumptions. The companion star of M2V class fills its Roche lobe. For the star we have taken into account the limb darkening (linear law with a darkening coefficient of $x = 0.62-0.63$ depending on the local surface temperature) and gravitational darkening effects ( $T\propto g^\beta$ with $\beta=0.08$, where $g$ is the local gravitational acceleration and $T$ is the local temperature on the star surface not affected by the X-ray heating, \citealt{Lucy1967}). The accretion (advection) disc around the relativistic object has a quasi-parabolic radial cross-section with a thickness increasing outwards described by the parameter $A$ (see \citealt{Khruzina2011} for details). The disc is elliptical, and the X-ray source is located at its focus. Also, we included in the model the region of interaction between the disc and the incoming gas stream in the form of a hot line and a hot spot \citep{Cherepashchuk2019A0620}. The sketch of the binary system is shown in Fig.~\ref{fig:scheme}. The synthetic light curves were obtained via the standard method described in \cite{WilsonDevinney1971}. Fluxes from elementary surface areas were calculated using the Planck function with the corresponding local temperatures. 

In general, our model has 20 free parameters. It is possible to derive all of them from observations only if a studied system is eclipsing which is not the case for \obj. Therefore, it was reasonable to decrease the number of free parameters by involving additional assumptions for the less important of them. The main goal of the modeling was to determine the orbital inclination $i$ of the system. The remaining free parameters were: two temperatures describing properties of the hot spot and hot line ($T_{\rm ww}$ and $T_{\rm lw}$, see below), the disc size $R_d$ (semi-major axis of the ellipse) and also the index $\alpha_g$ and the inner temperature $T_{\rm in}$ which define the temperature distribution over the disc surface:
\begin{equation}
T(r)=T_{\rm in} \left({R_{\rm in}\over r}\right)^{\alpha_g}.\
\label{eq_temp}
\end{equation}
The inner disc radius $R_{\rm in}$ as well as the the parameters described below were fixed. The index $\alpha_g$ was expected to be not far from 0.75 \citep{ShakSun1973}. 

The mass ratio of system components $q=M_x/M_v$ is known from the calculations based on rotational broadening of absorption lines in the donor star spectrum. There are two estimates: $q\approx9$ by \cite{Harlaftis1999Keck} who first measured the star rotational velocity $V_{\rm rot}\sin{i}=90_{-27}^{+22}$~km/s, and a more up-to-date $q\approx14$ by \cite{Petrov2017}. In the latter case the authors used the same value of the rotational velocity but employed a more advanced model of the star, which assumed that its shape is pear-like rather than spherical, and took into account the effects of gravitational and limb darkening. Thus, we considered the mass ratio by \cite{Petrov2017} as more reliable, nevertheless, $q=9$ has also been tested. For a particular choice of $q$, we calculated the size and shape of the donor star assuming that it fills its Roche lobe. The star temperature $T_2\cong 3200$~K was chosen to conform the spectral class M2V. Also we calculated two quantities used below as spatial scales: the binary separation $a_0$ and the distance $\xi$ from the compact object to the inner Lagrangian point L$_1$.

\begin{figure}
\includegraphics[width=0.48\textwidth]{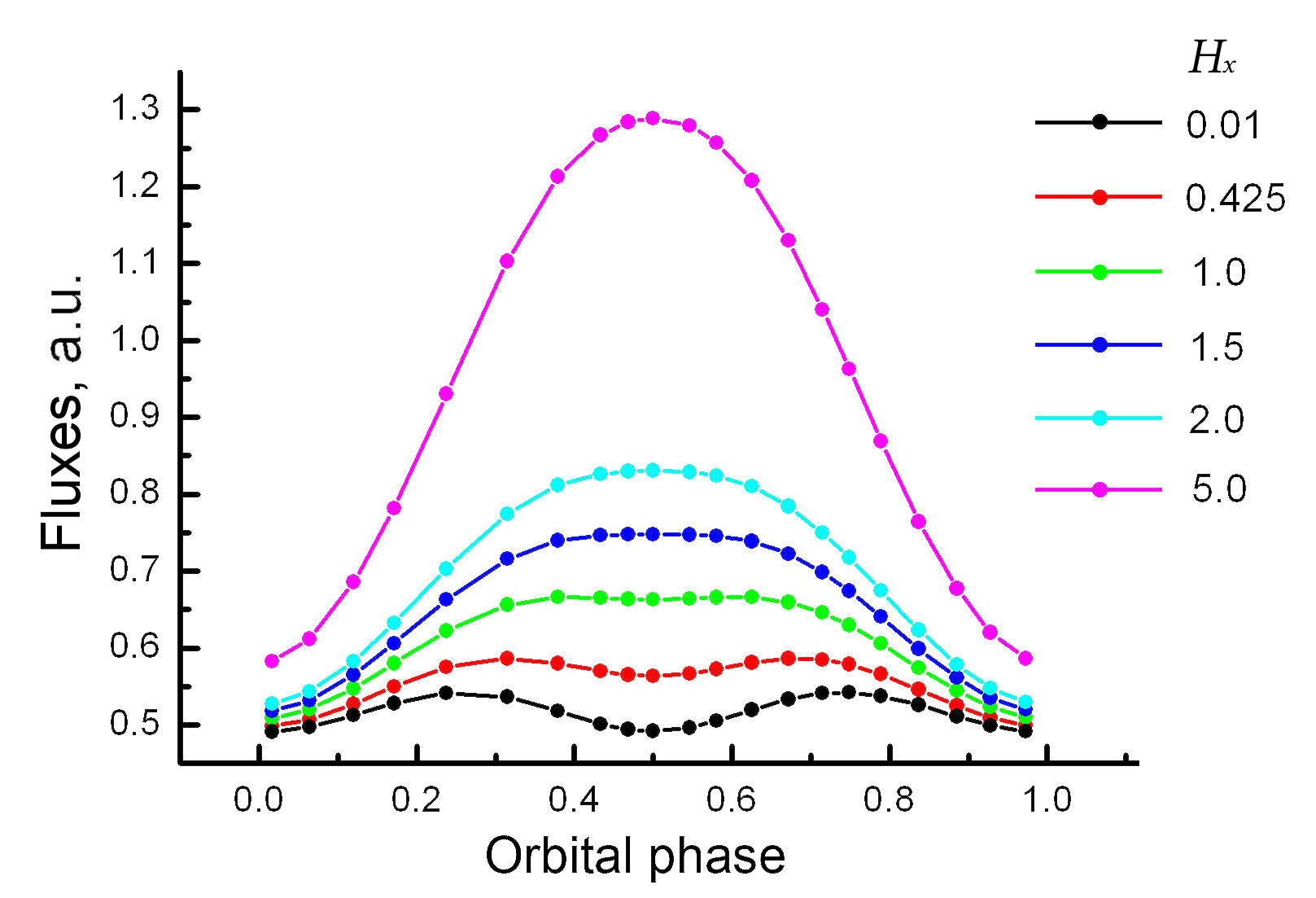}
\caption{Theoretical light curves for different luminosity ratios $H_x=L_x/L_v$. The effect of X-ray heating of the donor star become more prominent as $H_x$ increases.}
\label{fig:heating}
\end{figure}

X-ray novae demonstrate very low X-ray luminosities in their quiescent states, so the effect of heating of the donor star surface by X-rays coming from the central parts of the accretion disc is usually neglected when modelling their optical light curves (for example, \citealt{Cherepashchuk2019KVUMa}). However, as was mentioned above, in the case of \obj\ this effect may still be significant and has to be accounted. The importance of the heating effect is illustrated in Fig.~\ref{fig:heating} where we plot theoretical orbital light curves for different values of $H_x=L_x/L_v$. The curve for $H_x\approx0.4$ corresponding to $L_x = 4\times10^{31}$~erg/s still has an obvious minimum at $\varphi=0.5$ (the X-ray source is in front of the optical star). As $H_x$ increases, the double wave characteristic of the ellipticity effect is suppressed by the growing heating effect, which become completely dominating at $H_x>1.5$.

The X-ray heating of each elementary star surface area was taken into account by adding the bolometric flux from the not-heated star area and the irradiating bolometric flux from the X-ray source:
\begin{equation}
\sigma T^4 = \sigma T^4_0 + (1-\eta_s)F_x^{\rm bol},
\end{equation}
where $T$ and $T_0$ are the new and the initial temperatures of the star surface element, $\sigma$ is the Stefan-Boltzmann constant, $F_x^{\rm bol}$ is the incident X-ray flux and $\eta_s$ is albedo of the star surface (it is believed to be not higher than 0.5, \citealt*{deJong1996}).

\begin{figure*}
\includegraphics[width=0.98\textwidth]{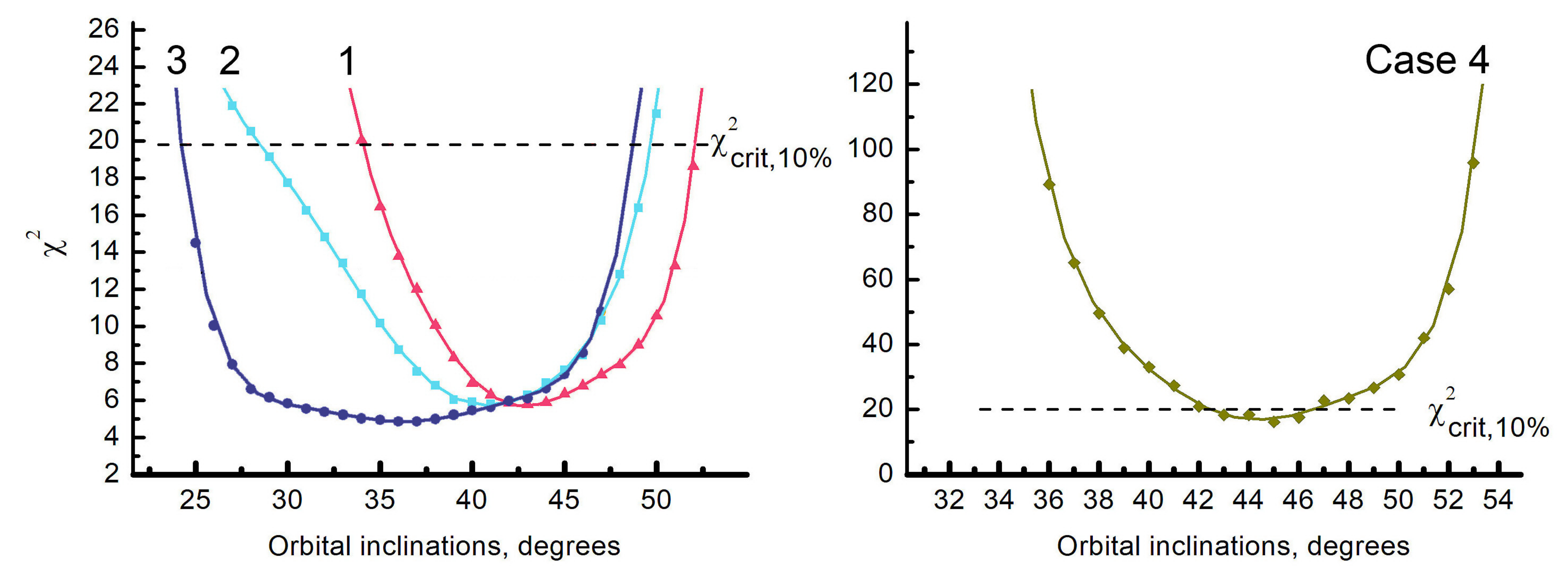}
\caption{Fit statistic as a function of the orbital inclination $i$ for models with different mass ratios $q$ and temperatures of the central source $T_1$. {\it Left panel}: $q=9$, $T_1 = 27\,720$~K (Case\,1, raspberry red); $q=9$, $T_1 = 60\,000$~K (Case\,2, cyan); $q=14$, $T_1 = 27\,720$~K (Case\,3, navy blue). {\it Right panel}: $q=14$, $T_1 = 40\,000$~K (Case\,4). Dashed line marks the 10\% critical level $\chi^2_{\rm crit, 10\%}=19.8$. The statistic was measured at discrete points with a step $\Delta i =1^\circ$, the curves are the approximations of these dependencies with a forth-order polynomial.}
\label{fig:chisq}
\end{figure*}

\begin{figure*}
\includegraphics[width=0.98\textwidth]{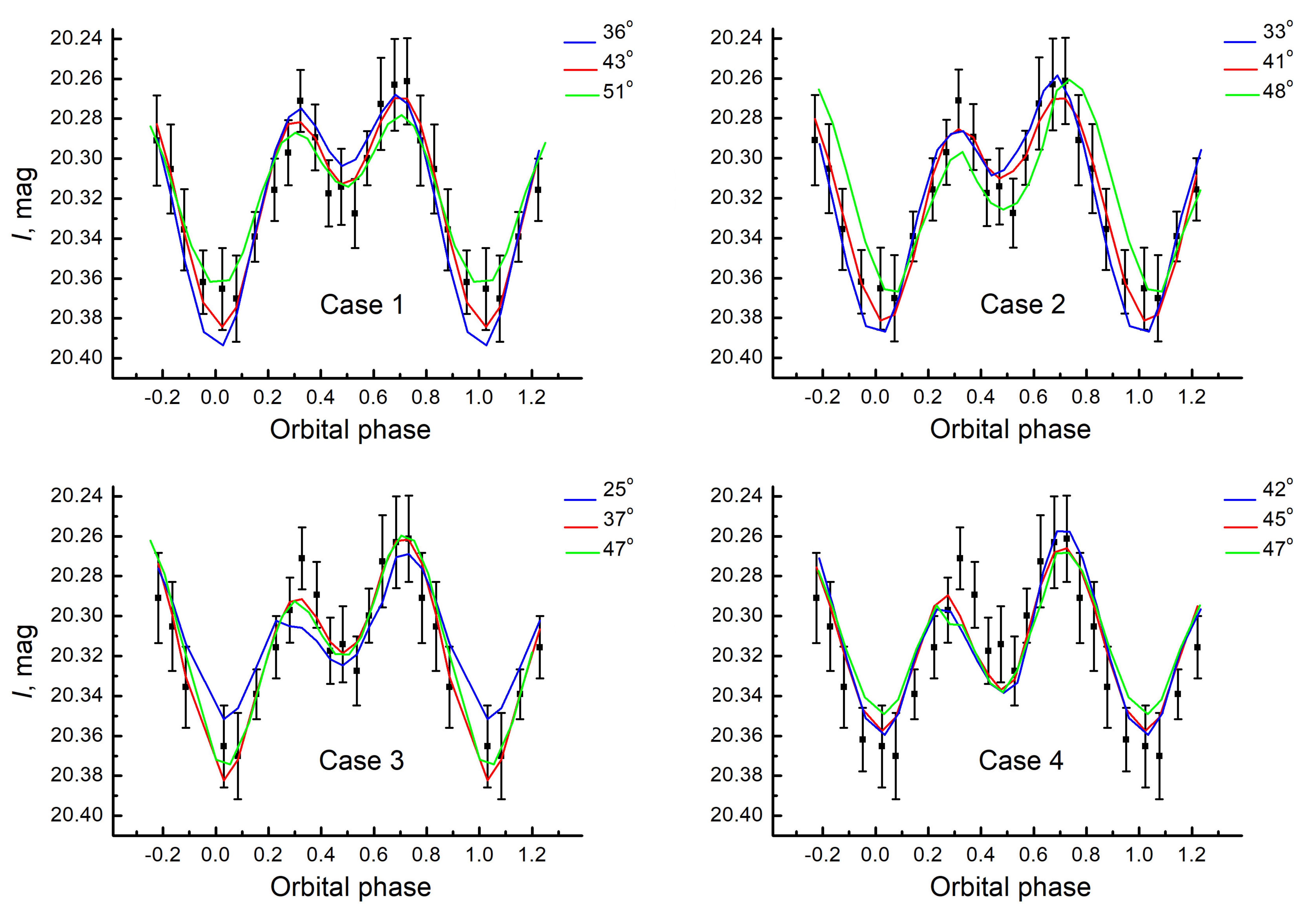}
%\begin{minipage}{\textwidth}
%\centering
%\includegraphics[width=0.42\textwidth]{fig06_q14T60.pdf}
%\hspace{2pt}
%\includegraphics[width=0.42\textwidth]{fig06_q9T60.pdf}
%\vspace{2mm}
%\end{minipage}
%\begin{minipage}{\textwidth}
%\centering
%\includegraphics[width=0.42\textwidth]{fig06_q14T40.pdf}
%\hspace{2pt}
%\includegraphics[width=0.42\textwidth]{fig06_q14T27.pdf}
%\end{minipage}
\caption{The observed orbital light curve together with the synthetic light curves for different considered temperatures $T_1$ and mass ratios $q$ (see Table~\ref{tab3:model} and the caption of Fig.~\ref{fig:chisq}). The red lines denote the models with the best-fitting values of the orbital inclination $i$ while the blue and green curves denote models with boundary values of $i$ corresponding to the critical level of the fit statistic.}
\label{fig:model_curves}
\end{figure*}

Due to the ambiguity of the physical models of central parts of an advection-dominated disc, the source of the heating emission was represented simply as a small sphere with a radius $R_1\equiv R_{\rm in}$ and a temperature $T_1$\footnote{We note that the temperature of the innermost disc ring $T_{\rm in}$ and the temperature of the `central source' $T_1$ are two different parameters. The spherical source was introduced because flat hot inner rings of the disc are much less efficient as an irradiating source (for the same temperature) because of their small surface area and orientation. Also, both $T_{\rm in}$ and $T_1$ were, strictly speaking, far below X-ray temperatures because of limitations of the computational grid, but this was not important for us, since we did not intend to model the X-ray emission. Only the bolometric luminosity was used to calculate the heating.} located in a focus of the (slightly) elliptical accretion disc. The temperature $T_1$ and radius $R_1$ are tied together with the bolometric luminosity of the X-ray source $L_x=L_{\rm bol}=4\pi\sigma T_1^4 R_1^2$ which should be close to typical X-ray luminosities in quiescence of X-ray nova with a black hole and a main sequence donor star ($L_x \sim 10^{30}\div 10^{32}$~erg~s$^{-1}$,  \citealt{Garcia2001,Cherepashchuk2013p2}), but possible values of $R_1$ was limited from below by the spatial resolution of our simulation grid. Thus we fixed the radius at $R_1 \cong 0.00062 \xi$ (i.\,e. $0.00046a_0$ for $q=14$ and $0.00044a_0$ for $q=9$) and used $T_1$ to set the luminosity ot the heating source. We tried $T_1 = 27\,720$\,K and 40\,000\,K for $q=14$ ($L_x = 3.2\times 10^{30}$ and $1.4\times 10^{31}$~erg~s$^{-1}$, respectively), as well as $T_1 = 27\,720$\,K and 60\,000\,K  for $q=9$ ($L_x = 2.2\times 10^{30}$ and $4.9\times 10^{31}$~erg~s). 

The eccentricity $e$ and orientation angle $\alpha_c$ of the disc (the angle between the direction to the disc periastron and the line connecting the centres of mass of the binary components) were fixed at $e \cong 0.01$ and $\alpha_c \cong 110^\circ$. The parameter defining the disc radial cross-section was $A=6.25$ \citep{Khruzina2011} which for the typical disc size $R_d/\xi \simeq 0.4 -0.5$ corresponds to the opening angle about $\beta_d\sim 1.5^\circ$. The hot line was represented by an truncated ellipsoid \citep{Khruzina2011} with semi-axes $a_v=0.040a_0$,  $b_v=0.244a_0$ and $c_v=0.0093a_0$. The relative size of the hot spot on the side surface of the disc $R_{\rm sp}/R_d \cong 0.1$. The temperature distribution over surfaces of the hot line was described by the parameters $T_{\rm ww}$ and $T_{\rm lw}$ \citep{Khruzina2011} which were free. Here indices `ww' and `lw' refer to the `windward' and `leeward' sides of the hot line facing the observer at the phases $\varphi=0.25$ and $\varphi=0.75$, respectively. The temperature of the hot spot was determined by the temperature of the hot line base.

\subsection{Modelling results}

\begin{table*}
\caption{Best-fit parameters.}
\begin{minipage}{12.3cm}
\begin{tabular}{lcccc}
\hline\hline
Parameter  &  Case\,1 & Case\,2 & Case\,3 & Case\,4  \\  \hline%
$q = M_x/M_v$  &  \multicolumn{2}{c}{\it 9.0} & \multicolumn{2}{c}{\it 14.0} \\
$T_1$,~K & {\it 27\,720} & {\it 60\,000} & {\it 27\,720} & {\it 40\,000} \\ 
$^*L_x$, erg/s & $2.2\times 10^{30}$ & $4.9\times 10^{31}$ & $3.2\times 10^{30}$ & $1.4\times 10^{31}$ \\

$^*\xi/a_0$   & 0.7090 & 0.7090 & 0.7434 & 0.7434 \\
$^*\langle R_v\rangle/a_0$   & 0.2198 & 0.2198 & 0.1932  & 0.1932 \\
$R_1/\xi$     & 0.00062 & 0.00062 & 0.00062 & 0.00062 \\
$T_v$, K      & 3\,200 & 3\,200 & 3\,200   & 3\,200 \\
$^*\beta_d, ^\circ$ & 1.5 & 1.5 & 1.5 & 1.5 \\
\hline
$i,^\circ$    & $43_{-2.8}^{+3.0}$  & $41_{-3.0}^{+2.8}$ & $36.5_{-6.6}^{+5.2}$ & $45_{-1.1}^{+0.9}$ \\ 
$R_d/ a_0$    & $0.417\pm0.033$ & $0.451\pm0.005$ & $0.439\pm0.14$ & $0.489\pm0.001$ \\
$^*R_d/\xi$   & $0.593\pm0.047$ & $0.643\pm0.007$ & $0.596\pm0.019$ & $0.664\pm0.001$ \\
$T_{\rm in}$, K & $127\,000 \pm 6\,000$ & $139\,000 \pm 5\,000$ & $119\,000 \pm 5\,000$ & $134\,000\pm 9\,000$\\
$\alpha_g$     & $0.641\pm0.013$ & $0.662\pm0.013$ & $0.667\pm0.012$ & $0.637\pm0.017$ \\
$^*T_{\rm out}$, K  & $1\,780\pm60$  & $1\,700\pm45$  & $1\,450\pm40$  & $1\,600\pm60$ \\
$T_{\rm ww}$, K   & $5\,600\pm1\,600$  & $9\,800\pm2\,600$ & $3\,700\pm1\,000$ & $10\,300\pm2\,900$ \\
$T_{\rm lw}$, K   & $5\,500\pm700$  & $9\,100\pm280$ & $5\,800\pm500$ & $10\,000\pm1\,500$ \\
$^*\langle T_{v, \rm heat}\rangle$, K & 3\,233 & 3\,265 & 3\,215 & 3\,250 \\
$^*F_{\rm star}/F_{\rm sys}$ & $0.494\pm0.009$ & $0.497\pm0.011$ & $0.611\pm0.008$ & $0.351\pm0.008$ \\
$^*F_{\rm disc}/F_{\rm sys}$ & $0.458\pm0.001$ & $0.394\pm0.002$ & $0.339\pm0.001$ & $0.557\pm0.001$ \\
$^*F_{\rm HL}/F_{\rm sys} $ & $0.048\pm0.002$ & $0.108\pm0.004$ & $0.050\pm0.004$ & $0.092\pm0.006$ \\
$^*F_{\rm CO}/F_{\rm sys}$ & 0.0003 & 0.0008 & 0.0006 & 0.0005 \\ 
$\chi^2_{\rm min}$  & 5.783 & 5.791 & 4.863 & 16.121 \\
 \hline
\end{tabular}
{\it Notes.} The top part of the table lists the most important of the parameters fixed based on additional information about the system. We have considered four options of the same model defined by pairs of the mass ratio $q$ and temperature $T_1$. The temperature $T_1$ together with the radius $R_1$ set the bolometric luminosity $L_x$ of the compact object at the disc centre. The temperature of the donor star without heating $T_v$ corresponds to its spectral class M2V. The mean size of the star filling its Roche lobe $\langle R_v\rangle$ and the distance $\xi$ from the disc centre to the Lagrangian point L1 computed for the particular $q$ are given in units of the binary separation $a_0$. $\beta_d$ is the approximate opening angle of the accretion disc. The bottom part shows 6 desired parameters obtained from the modelling with their 1$\sigma$ errors: the orbital inclination $i$, the mean radius of the elliptical (in general case, Sec.\,\ref{sec:model_intro}) accretion disc $R_d$, the inner temperature $T_{\rm in}$ and the index $\alpha_g$ describing the temperature distribution in the disc (eq.\,\ref{eq_temp}), the windward $T_{\rm ww}$ and leeward $T_{\rm lw}$ temperatures of the hot line. Asterisks mark the parameters dependent on the others (the fixed or the desired), they were calculated in the process of solving the problem. The mean temperature at the disc outer edge $T_{\rm out}$ is obtained using eq.\,(\ref{eq_temp}), $\langle T_{v, \rm heat}\rangle$ is the mean temperature of the donor star after accounting for the heating. $F_{\rm star}$, $F_{\rm CO}$, $F_{\rm disc}$ and $F_{\rm HL}$ are fluxes from the donor star, compact object, accretion disc and hot line, respectively, in the units of the total system flux $F_{\rm sys}$.
\end{minipage}
\label{tab3:model}
\end{table*}

In total, we have tested four cases: $q=9$ with $T_1 = 27\,720$~K (Case\,1) and 60\,000~K (Case\,2), $q=14$ with $T_1 = 27\,720$~K (Case\,3) and 40\,000~K (Case\,4). Six parameters were free: $i$, $R_d$, $T_{\rm in}$, $\alpha_g$, $T_{\rm ww}$ and $T_{\rm lw}$. The solution of the inverse problem was searched for by minimization of the functional $\chi^2$~--- the weighted sum of squared deviations of the theoretical light curve from the observed one~--- using the Nelder-Mead optimization method \citep{Himmelblau1972}. To assess the sensitivity of the model to the orbital inclination $i$ as well as to determine the uncertainties of this parameter, we have calculated $\chi^2$ as a function of $i$. To do this, we took some value of $i$ near its best-fitting value, fixed it and performed minimization of the functional over the remaining free parameters. Then we took another value of $i$ and repeated this procedure until the desired range of $i$ was entirely scanned. The step was $\Delta i = 1^\circ$. Since the parameter space has a complicated shape with numerous local minima, we employed the technique described by \cite{Charbonneau1995} to select the parameter values belonging to the same family. This allowed us to obtain smooth dependence of the relative luminosities of the binary components on the orbital inclination. To make sure that the found minimum of the functional is not local, we repeated the optimization from many different initial guesses (several dozens).

The obtained fit statistic $\chi^2$ as a function of the inclination for each of the considered pairs of $T_1$ and $q$ is shown in Fig.~\ref{fig:chisq}. Cases~1--3 gave a good agreement of the model with the observed orbital light curve resulting in $\chi^2_{\rm min}\approx 4.9 - 5.8$ (Fig.~\ref{fig:chisq}a) while the critical level corresponding to a significance of $\alpha=10\%$ is $\chi^2_{\rm crit} = 19.8$ (for 13 degrees of freedom\footnote{Because the functional minimum was searched for by iterating over $i$, only 5 parameters remained free. The number of data point in the binned light curve (Table~\ref{tab2:binned}) is 18.}). The degree of the model sensitivity to different values of $i$ can also be assessed from Fig.~\ref{fig:model_curves} where we plot the synthetic light curves corresponding to the best-fitting and two subsidiary values of $i$ together with the observational data. 
In Case\,4, we were unable to obtain fit statistic better than $\chi^2\simeq 16.1$ (Fig.~\ref{fig:chisq}b). Also, this model has shown a different behaviour: it is the only one of the four considered in which the optical emission is clearly dominated by the contribution of the accretion disk rather than the donor star (see below). The poor agreement of this model ($q=14$, $T_1=40\,000$~K) is the reason why we did not try to raise the temperature $T_1$ any higher (say up to 60 000~K). 

For Cases~1--3, the fit statistic normalized to the number of degrees of freedom is appeared to be less than unity. This is generally considered to be not good because a model starts to fit not only the regular signal, but also the noise. However, our observed orbital light curve is very smooth, and the model curves pass almost perfectly through its midpoints: practically none of the 18 data points of the observed curve deviates more than one root-mean-square error  from the model curves. Thus, our results can be considered satisfactory.

The best-fitting model parameters are given in Table~\ref{tab3:model}. In order to measure the parameter uncertainties, we employed the $\chi^2_{\rm min}+1$ criterion which gives a 1$\sigma$ (68.3\%) confidence interval for a single `parameter of interest'. Inclination of the system is turned out to be close to 40$^\circ$ in all the cases. The disc inner temperature $T_{\rm in}$ lies in the range from $\simeq 119\,000$~K to $\simeq 139\,000$~K, the  size of the disc is about 60\% of the compact object's Roche lobe. The outer temperature is only $\simeq 1\,500$~K, which is by a factor of 3.5--6 lower compared to the temperature of the hot line.

\begin{figure*}
\includegraphics[width=0.98\textwidth]{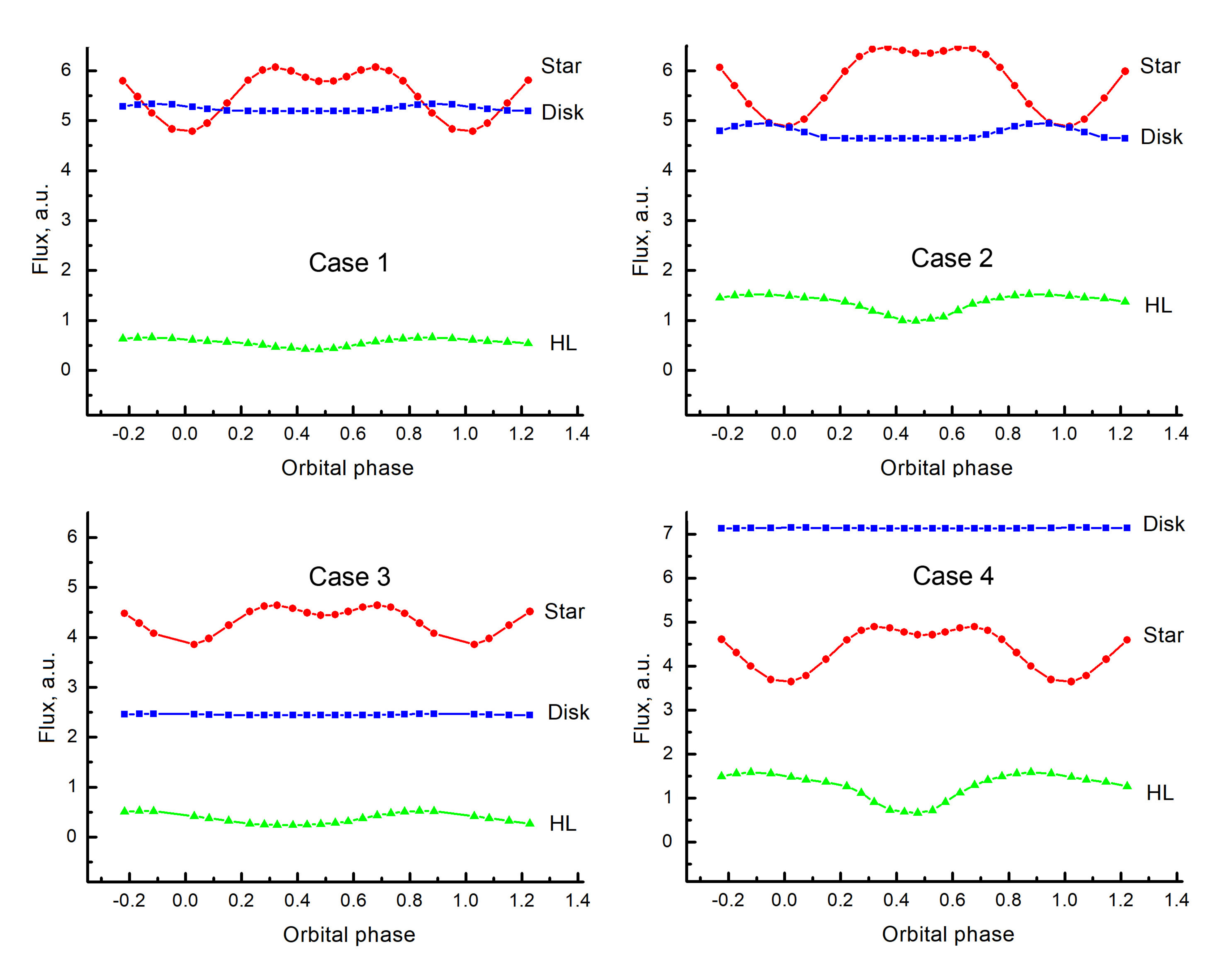}
\caption{Model light curves decomposed into contributions of individual system components: the accretion disc (blue), the donor star (red) and the hot line/hot spot (green) for the four considered cases. The fluxes are in arbitrary units.}
\label{fig:components}
\end{figure*}

In order to grasp the model behaviour, in Fig.~\ref{fig:components} we plot synthetic light curves of individual system components. The double-humped light curve is described by the ellipticity effect of the donor star. The depth of its secondary minimum determined by the strength of the heating effect turned out to be weakly dependent on the luminosity $L_x$. This is because the accretion disc also heats the donor star, especially its middle rings, which are still not too cold but have relatively large surface area. The orbital curve of the disc is almost a straight line. Its slight curvature is determined by the visibility conditions of the hot spot on the disc side surface. The hot line, due to the difference between temperatures of its windward and leeward sides, produces an asymmetric light curve whose minimum is shifted respect to the secondary minimum of the donor star's curve. This asymmetry allowed to describe the observed difference in the heights of the humps in the total light curve.

Both cases of $q=9$ yielded almost equal contribution of the stellar and non-stellar components which is consistent with the results of the spectroscopic observations \citep{Casares1995,Harlaftis1999Keck}. In flux units, the star is slightly brighter in the cases with $q=9$ compared to $q=14$ due to the larger Roche lobe (in units of the binary separation). The relative contribution of the donor star in the cases with $q=14$ appeared to be much different (Table~\ref{tab3:model}, Fig.~\ref{fig:components}). Thus, the model with higher $T_1$ (Case\,4) gave hotter, larger, and, therefore, more luminous accretion disc with the same luminosity of the companion star, which decreased the star contribution from 61\% (Case\,3) to 35\% (Case\,4). Since distance to the system is poorly known, we unable to distinguish which option takes place in reality. In principal, the disc inner temperature can be determined from UV observations. As we have estimated from our model, the disc has to be fainter at $\sim3000\AA$ by about two stellar magnitudes than it is in the I band. We note that although the Case 4 showed worse agreement with the observed data, this model has a formal significance of 0.24 (p-value) and still can not be rejected.
  
%{\bf The main goal of this modeling was to constrain possible values of the orbital inclination $i$ which is important when measuring masses of binary components. To impose such constraints in the most conservative manner, we have used the critical values shows by dashed lines in Fig.~\ref{fig:chisq}. We note that the ranges determined in such a way are not uncertainties in the usual sense, they are projections of the five-dimensional confidence area onto the axis of the desired parameters \citep{Cherepashchuk2022ARep66S70S122} and represent some boundaries within which the model is still not too bad. Since the Cases~1--3 fit the observed curve much better, we decided to choose critical values corresponding to the confidence level $\alpha=0.5$ for them. For Case\,4, the confidence level is chosen to be $\alpha=0.1$. The resultant ranges of $i$ are shown in the bottom part of Table~\ref{tab3:model}.}

\section{Masses of the binary components}
The main result of this work is the estimates of the orbital inclination $i$, which allows to calculate masses of the binary components using known values of the mass ratio $q$ and mass function of the donor star. We used the value of the mass function  $f_v(M_x)=1.21\pm 0.06$~M$_\odot$ reported by \cite{Filippenko1995}. From the definition of the mass function of the secondary 
$$f_v(M_x)=\frac{M_x^3\sin^3 i}{(M_x+M_v)^2},$$ the mass the compact object can be expressed as:
\begin{equation}
M_x= \frac{f_v(M_x)(1+q)^2}{q^2\sin^3 i},
\end{equation}
and $M_v = M_x/q$.
The resultant masses calculated for each of the four considered cases are listed in Table~\ref{tab4:masses} where we have taken into account uncertainties of both the measured inclination and the mas function. The total mass ranges are $M_x=(3.9-6.3)M_\odot$, $M_v=(0.43-0.7)M_\odot$ for $q=9$ and $M_x=(3.6-9.4)M_\odot$, $M_v=(0.26-0.67)M_\odot$ for $q=14$.

%\begin{table}
%\caption{Masses of the binary components computed for the obtained $i$.}
%\centering
%\begin{tabular}{cccc}
%\hline\hline
% &~~ $i,^\circ$ ~~& $M_x$, $M_\odot$ & $M_v$, $M_\odot$ \\
%\hline
%\multirow{3}{*}{\makecell{Case\,1\\($q=9$, $T_1=27\,000$\,K)}}
% & 43 & $4.71\pm0.23$ & $0.523\pm0.026$ \\
% & 37 & $6.85\pm0.34$ & $0.762\pm0.038$  \\
% & 45 & $4.23\pm0.21$ & $0.470\pm0.024$ \\ \hline
%\multirow{3}{*}{\makecell{Case\,2\\($q=9$, $T_1=60\,000$\,K)}}
% & 41 & $5.29\pm0.26$ & $0.588\pm0.029$ \\
% & 37 & $6.85\pm0.34$ & $0.762\pm0.038$ \\
% & 44 & $4.46\pm0.22$ & $0.495\pm0.024$ \\ \hline
%\multirow{3}{*}{\makecell{Case\,3\\($q=14$, $T_1=27\,000$\,K)}}
% & 37 & $6.37\pm0.32$ & $0.455\pm0.022$ \\
% & 33 & $8.60\pm0.43$ & $0.614\pm0.030$ \\
% & 39 & $5.57\pm0.28$ & $0.398\pm0.020$ \\ \hline
%\multirow{3}{*}{\makecell{Case\,4\\($q=14$, $T_1=40\,000$\,K)}}
% & 45 & $3.93\pm0.20$ & $0.281\pm0.014$ \\
% & 42 & $4.64\pm0.23$ & $0.331\pm0.016$ \\
% & 49 & $3.23\pm0.16$ & $0.231\pm0.012$ \\ \hline
%\end{tabular}
%\label{tab4:masses}
%\end{table}

\begin{table}
\caption{Masses of the binary components computed for the obtained $i$.}
\centering
\begin{tabular}{llcc}
\hline\hline
 &~~~$i,^\circ$ & $M_x$, $M_\odot$ & $M_v$, $M_\odot$ \\
\hline
Case\,1 ($q=9$, $27$~kK) & $43\pm 2.9$ & $4.7\pm0.8$ & $0.52\pm0.09$ \\
Case\,2 ($q=9$, $60$~kK) & $41\pm2.9$ & $5.3\pm1.0$ & $0.59\pm0.11$ \\
Case\,3 ($q=14$, $27$~kK) & $36.5\pm5.9$ & $6.6\pm2.8$ & $0.47\pm0.20$ \\
Case\,4 ($q=14$, $40$~kK) & $45\pm1.0$ & $3.93\pm0.29$ & $0.281\pm0.020$ \\ \hline
\end{tabular}
\label{tab4:masses}
\end{table}

We emphasize that the uncertainties of the determined masses stem not only from observational errors, but also from ambiguity of the \obj\ system state related to the fact that we do not have X-ray observations  synchronous with our optical data. As a result we had to vary the parameters governing the heating effect in a wide range in order to take into account the possible physical variability of the system in its quiescent state.

The cases with $q=9$ yielded the donor mass $(0.43-0.7)M_\odot$ which is thought to be too high for a M2V star ($\sim 0.4M_\odot$). Therefore the results for $q=14$ look more preferable. Thus, as the final result we accept values $M_x=(6.5\pm2.9)M_\odot$ and $M_v=(0.47\pm0.21)M_\odot$. The determined range of the possible values of the black hole mass overlaps with the known 2--5\,M$_\odot$ gap in the distribution of compact objects masses but mostly it lies above the upper boundary of this gap.

%We emphasize that the obtained ranges are the most conservative estimates of the masses $M_x$ and $M_v$. If we take only the optimal values of $i$ then the mass range become narrower, however, this yields $M_v=(0.52-0.58)$~M$_\odot$ for $q=9$, which looks too low for the M2V donor star. Thus, $q=14$ with $M_v=(0.28-0.46)$~M$_\odot$ and $M_x = (3.9-6.4)$~M$_\odot$ should be considered as more reliable. \blue{This imply that smaller estimates from the determined range of the black hole mass may still fall into the known 2--5\,M$_\odot$ gap in the mass distribution of compact objects but lie close to its upper bound.}}

\section{Discussion and Conclusions}
We have analysed new photometric observations of the quiescent X-ray nova \obj\ (V518\,Per) performed in the I$_c$ band in 2020--2023 for five epochs. We were unable to obtain a satisfactory orbital light curve using the ephemeris by \cite{Webb2000} based on the spectroscopy carried out more than 20 years before our observations. In order to build a regular average orbital light curve from our observations, we had to significantly revise the orbital period of the system. Our new value is  $P_{\rm orb} = 5^{\rm h}04^{\rm m}35\fs50\pm0\fs04$ which is statistically consistent with estimates by \cite{Filippenko1995} ($5^{\rm h}04^{\rm m}41^{\rm s} \pm 49^{\rm s}$) and \cite{Harlaftis1999Keck} ($5^{\rm h}04^{\rm m}43^{\rm s} \pm 29^{\rm s}$).

The orbital light curve was simulated within the model of an interacting binary system, which, among other things, takes into account the presence of the hot line near the outer boundary of the disk, i.\,e. the gas flow that collides with the rotating outer parts of the disk. As a result, we constrained possible values of the orbital inclination of the system $i$ and calculated corresponding estimated for masses of both the black hole and the companion star using the mass function from \cite{Filippenko1995}. Our value of the black hole mass $M_x=(6.5\pm2.9)M_\odot$ lies between the estimates $(3.97\pm0.95)M_\odot$ by \cite{GelinoHarrison2003} and $M_x\gtrsim9M_\odot$ by \cite{Beekman1997} and \cite{Reynolds2007Keck}. The relatively large uncertainties of our estimate stem from the fact that we were unaware about X-ray heating conditions of the system at the moments of our optical observations and had to try different options. So to obtain a more precise result one needs to conduct joint X-ray and optical observations to constrain luminosity of the X-ray source.

The confidence interval of our estimate partially overlaps with the known 2--5\,M$_\odot$ gap in the mass distribution of compact objects. The nature this gap remains unclear. It is only known that there is a strong deficit of low-mass black holes, and this conclusion is supported by extensive observational data. In particular, the gravitational wave channel afforded mass measurements for more than one and a half hundred black holes and several neutron stars \citep{Abbott2023}. Most of these black holes are heavy, the only known lightweight specimen falling into the gap is GW230529 of $(2.5-4.5)M_\odot$ discovered very recently \citep{LIGO2024arXiv240404248T}. In the electromagnetic channel, the number of measurements of neutron star masses is currently approaching 100, the number of the measured black holes masses exceeds 30. Among these black holes, \obj\ is one of the lightest.

Such a distribution of masses suggests that the mass of a newborn compact object forming during core collapse may depend not only on the mass of its progenitor but also on additional stellar parameters, such as the magnetic field of the progenitor's core, its rotation, some instabilities that can arise during the collapse and determine the statistical outcome of the collapse, and so on (see, for example, \citealt{Ergma1998}). This makes it important to measure the black hole mass in \obj\ as precise as possible, so further observations of this system is quite perspective.

%The most conservative approach to the error estimation gave a wide range of \chk{$i$ from xxx to xxx}, which corresponds to the mass ranges $M_x=(3-19)$~M$_\odot$ and  $M_v=(0.2-1.2)$~M$_\odot$. Taking only the optimal values of $i$, at which the minimum of the fit statistic is reached, and bearing in mind the spectral type of the donor, the mass ranges could be shrunk to $M_v=(0.28-0.46)$~M$_\odot$ and $M_x = (3.9-6.4)$~M$_\odot$. In this case the mass of the donor becomes consistent with its spectral type M2V, and the range of possible values of the black hole mass intersects with the 2--5\,M$_\odot$ gap in the mass distribution of compact objects. Thus, our result is consistent with the conclusions by \cite{GelinoHarrison2003} that the \obj\ system has a relatively lightweight black hole which potentially can fall into the mass gap.

\section*{Acknowledgements}
The authors would like to thank the anonymous referees for useful comments on this paper. AMCh acknowledge support from the Russian Science Foundation grant 23-12-00092 (formulation of the problem, interpretation of observations). 

\section*{Data Availability}
The data underlying this article will be shared on reasonable request to the corresponding author. 

\bibliographystyle{mnras}
\bibliography{bibtexbase.bib}

\appendix
\section{Modeling for the less probable period $P_{\rm orb}=0\fd21216275$}

As was mentioned above, the observations being folded with the ephemeris (\ref{eq:eph_near_webb}) produce an orbital light curve with no minimum at $\varphi=0.5$ (Fig.~\ref{fig:lcurve}b) which can be caused by significant heating in the system. We have tested two cases: $T_1=27\,000$~K (Case\,A1, small heating luminosity $L_x = 3.2\times 10^{30}$~erg~s$^{-1}$) and $T_1=60\,000$~K (Case\,A2, $L_x = 7\times 10^{31}$~erg~s$^{-1}$), both for $q=14$. The remaining model parameters were the same as above.

\begin{figure}
\includegraphics[width=0.48\textwidth]{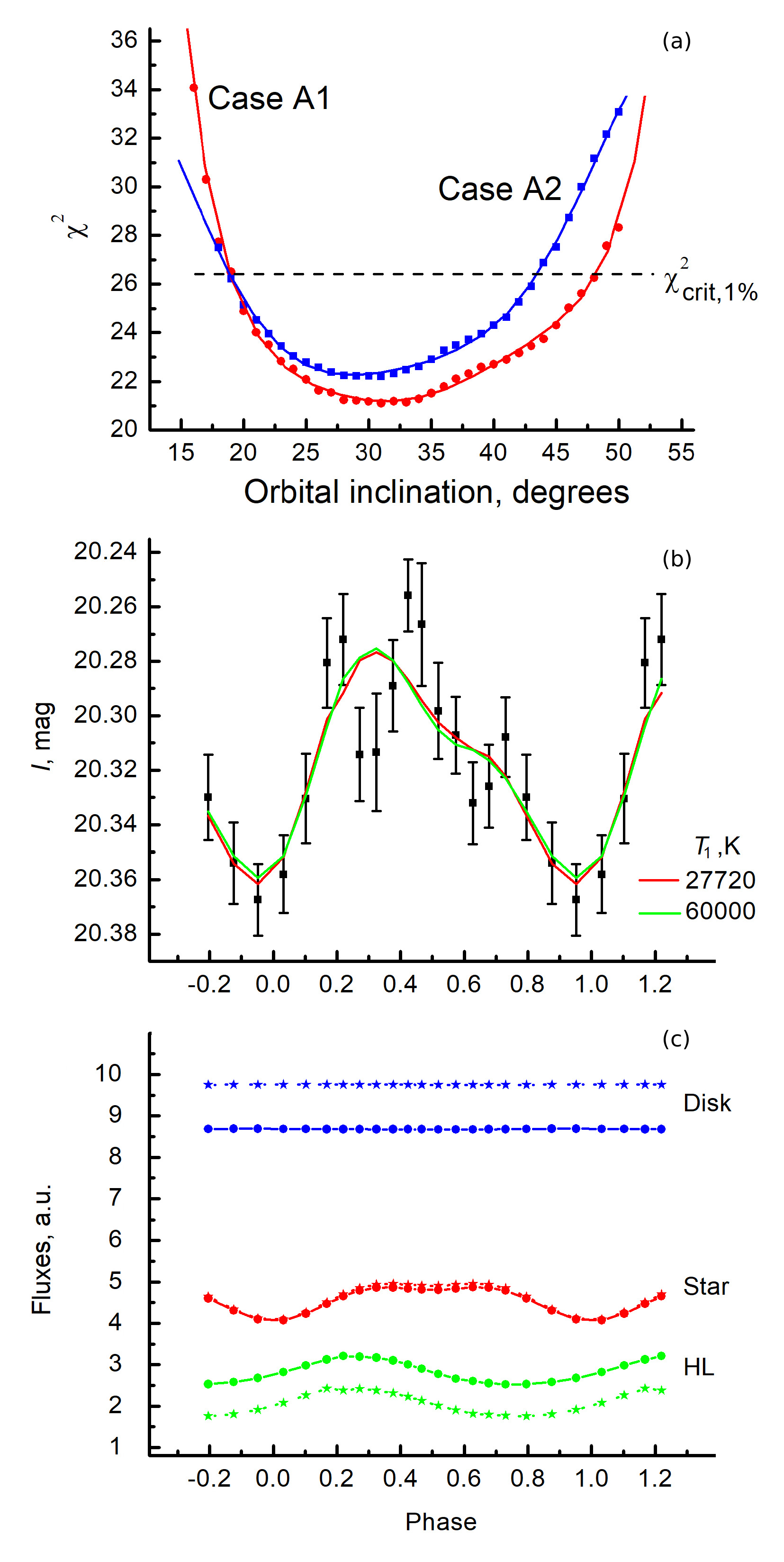}
\caption{{\it Top panel:}  Fit statistic as a function of the orbital inclination for $q=14$, $T_1 = 27\,720$~K (Case\,A1) and $q=14$, $T_1 = 60\,000$~K (Case\,A2). {\it Middle panel:} Observed and synthetic light curves for the same models. {\it Bottom panel}: model light curves decomposed into contributions of individual system components: the accretion disc (blue), the donor star (red) and the hot line/hot spot (green). The model with $T_1=60\,000$~K is shown with circles and the solid line, the model with  $T_1=27\,720$~K with stars and the dashed line. The fluxes are in arbitrary units.}
\label{fig:appendix}
\end{figure}

The result is shown in Fig.~\ref{fig:appendix} and Table~\ref{tab:modelA}. It is seen that, qualitatively, both models have been able to reproducing such a peculiar light curve but the goodness of these fits is poor. The reached minima of the statistic correspond to the significance\footnote{For 12 degrees of freedom since the folded light curve now contains 17 bins.} 0.049 and 0.035 for Case\,A1 and Case\,A2, respectively.

\begin{table}
\caption{Best-fit model parameters for $P_{\rm orb}=0\fd21216275$.}
\begin{minipage}{8.5cm}
\begin{tabular}{lcc}
\hline\hline
Parameter  &  Case A1 & Case A2 \\  \hline%
$q = M_x/M_v$  & \multicolumn{2}{c}{\it 14.0} \\
$T_1$,~K & {\it 27\,720} & {\it 60\,000} \\ 
$^*L_x$, erg/s & $3.2\times 10^{30}$ & $7.0\times 10^{31}$ \\ 
\dots & \dots  & \dots \\
\hline
$i,^\circ$  & $31_{-5.9}^{+6.3}$ & $31_{-7.3}^{+5.2}$ \\ 
$R_d/ a_0$  & $0.343\pm0.002$ & $0.347\pm0.006$  \\
$^*R_d/\xi$ & $0.465\pm0.002$ & $0.472\pm0.009$  \\
$T_{\rm in}$, K  & $132\,000\pm13\,000$ & $128\,000\pm14\,000$  \\
$\alpha_g$       & $0.645\pm0.011$ & $0.644\pm0.012$  \\
$^*T_{out}$, K   & $1\,870\pm160$ & $1\,880\pm180$  \\
$T_{\rm ww}$, K  & $9\,290\pm1\,100$ & $9\,800\pm1\,200$  \\
$T_{\rm lw}$, K  & $5\,870\pm380$ & $6\,800\pm600$  \\
$^*\langle T_{v, \rm heat}\rangle$, K & 3\,244 & 3\,236  \\
$^*F_{\rm star}/F_{\rm sys}$ & 0.284 & 0.287  \\
$^*F_{\rm disc}/F_{\rm sys}$ & 0.590 & 0.537  \\
$^*F_{\rm HL}/F_{\rm sys} $ & 0.126 & 0.176  \\
$^*F_{\rm BH}/F_{\rm sys}$ & 0.0003 & 0.0007  \\
$\chi^2_{\rm min}$ & 21.104 & 22.206  \\
 \hline
\end{tabular}
\end{minipage}
{\it Notes.} See notes under Table~\ref{tab3:model} for parameter description. Ellipsis denotes the fixed parameters that remained the same as in Table~\ref{tab3:model}.
\label{tab:modelA}
\end{table}

Both models yielded very similar best-fitting parameters with the same orbital inclination $i=31^\circ$ (Table~\ref{tab:modelA}). This inclination corresponds to the component masses $M_x=10.2M_\odot$ and $M_v = 0.73M_\odot$. The main difference of these models from the models considered in the main part of the paper (Table~\ref{tab3:model}) is much smaller contribution of the donor star to the total flux which has become here less than 30\%. The shapes of the total synthetic light curves (Fig.~\ref{fig:appendix}b) demonstrate very small differences. In the model with lower $T_1$, the disc appeared to be hotter and brighter (blue dashed line in Fig.~\ref{fig:appendix}c) which has compensated the lack of the luminosity $L_x$. As a result, the light curves of the donor star almost coincide in the figure. At the same time, the hot line/hot spot became colder in order to preserve the share of the non-stellar emission.

Contribution of the hot line has increased up to $\simeq 15\%$ compared to the results in Table~\ref{tab3:model}. It is the presence of a relatively bright hot line allowed to satisfactorily reproduce the complex asymmetric shape of the observed light curve. Such a result does not look too strange and can be explained by increased accretion rate. A rise of accretion rate is possible, because the donor star is of a late spectral class and has active regions with strong magnetic fields on its surface. When one of the active regions approaches the Lagrange point, this can change the disc mass inflow rate. Therefore, both the accretion disk and the hot line are `breathing'.

\end{document}